%% file: tcpcpt.tex
\documentclass[article,12pt,sort&compress]{elsarticle}

\usepackage{graphicx}
\usepackage{xspace}
\usepackage{amsmath}
\usepackage{amsfonts}
\usepackage{amssymb}
\usepackage{enumerate}
\usepackage{epsfig}
\usepackage{changepage}
\usepackage{lineno}%line number
\newcommand{\ksn}{\ensuremath{\mathrm{K_S}}\xspace}
\newcommand{\ksln}{\ensuremath{\mathrm{K_{S,L}}}\xspace}
\newcommand{\kln}{\ensuremath{\mathrm{K_L}}\xspace}

\newcommand{\kn}{\ensuremath{\mathrm{K^0}}\xspace}
\newcommand{\knb}{\ensuremath{\mathrm{\bar{K}{}^0}}\xspace}

\newcommand{\kpp}{\ensuremath{\mathrm{K_+}}\xspace}
\newcommand{\kppnn}{\ensuremath{\mathrm{K_{\pm}}}\xspace}
\newcommand{\knn}{\ensuremath{\mathrm{K_-}}\xspace}
\newcommand{\kpport}{\ensuremath{\mathrm{K_+^{\perp}}}\xspace}
\newcommand{\knnpport}{\ensuremath{\mathrm{K_{\mp}^{\perp}}}\xspace}
\newcommand{\knnort}{\ensuremath{\mathrm{K_-^{\perp}}}\xspace}

\newcommand{\bn}{\ensuremath{\mathrm{B^0}}\xspace}
\newcommand{\bbn}{\ensuremath{\mathrm{B}}\xspace}

\newcommand{\bnb}{\ensuremath{\mathrm{\bar{B}^0}}\xspace}

\renewcommand{\bar}{\overline}
\newcommand{\kaon}{\mathrm{K}^0}
\newcommand{\akaon}{\bar{\mathrm{K}}^0}
\newcommand{\Ks}{\mathrm{K_S}}
\newcommand{\Kl}{\mathrm{K_L}}

\newcommand{\Km}{\mathrm{K_{-}}}

\usepackage{subcaption}
\usepackage{amsmath}
\usepackage{soul}
\def\CP       {\ensuremath{\mathrm{CP}}\xspace}
\def\CPT       {\ensuremath{\mathrm{CPT}}\xspace}
\def\C       {\ensuremath{\mathrm{C}}\xspace}
\def\P       {\ensuremath{\mathrm{P}}\xspace}
\def\T       {\ensuremath{\mathrm{T}}\xspace}
\def\SS    {\ensuremath{\mathrm{s}}\xspace}
\newcommand{\affuni}[2]{Dipartimento di Fisica dell'Universit\`a #1, #2, Italy.}
\newcommand{\affinfn}[2]{INFN Sezione di #1, #2, Italy.}

\journal{Physics Letters B}

\begin{document}
\begin{frontmatter}
\title{Direct tests of \T, \CP, \CPT symmetries in transitions of neutral K mesons with the KLOE experiment}

\input{authors.tex}

\begin{abstract}
%A study of the $\Delta t$ distributions of
%%studying 
%the $\phi\rightarrow\ksn\kln\rightarrow\pi^+\pi^- \, \pi e \nu $ and 
%$\phi\rightarrow\ksn\kln\rightarrow \pi e \nu \, 3\pi^0$ processes,
%%decay channels and their $\Delta t$ distributions 
%with $\Delta t$ being the difference of the kaon decay times, is perfo are studied
%
%A data sample collected by the KLOE experiment at 
%DA$\Phi$NE, the Frascati $\phi$-factory, corresponding to 
%%$1.7 \times 10^9$
%%%billion 
%%$\phi\rightarrow\ksnkln$
%%decays produced
%an integrated
%luminosity of about 1.7 fb$^{-1}$ 
%is analysed to study 
%the $\Delta t$ distributions of
%%studying 
%the $\phi\rightarrow\ksn\kln\rightarrow\pi^+\pi^- \, \pi e \nu $ and 
%$\phi\rightarrow\ksn\kln\rightarrow \pi e \nu \, 3\pi^0$ processes,
%%decay channels and their $\Delta t$ distributions 
%with $\Delta t$ the difference of the kaon decay times.
%%% Text of abstract
\par
%A 
%direct 
Tests of the \T, \CP and \CPT symmetries in the neutral kaon system 
%has been
are performed 
%for the first time 
by 
%comparing 
%
%on 
the direct comparison of 
the probabilities of a
%flavor$\leftrightarrow$\CP state transition
kaon transition process
%a kaon transition  
%$i \rightarrow f$ 
%with
to 
%and
its symmetry-conjugate.
%d process.
%A new kind of \CPT test for transitions in the neutral kaon system is presented, where 
The exchange of 
{\it in} and {\it out} states 
%(and \CP conjugation), 
required for a genuine 
%\T  or \CPT 
test 
involving an anti-unitary transformation implied by time-reversal 
%\T
%under time-reversal 
%\T 
is
%can be 
%performed 
implemented
exploiting the 
%maximal 
entanglement of $\kn\knb$ pairs produced at a $\phi$-factory.
\par
A data sample collected by the KLOE experiment at 
DA$\Phi$NE corresponding to 
%$1.7 \times 10^9$
%%billion 
%$\phi\rightarrow\ksnkln$
%decays produced
an integrated
luminosity of about 1.7 fb$^{-1}$
is analysed to study 
the $\Delta t$ distributions of
%studying 
the $\phi\rightarrow\ksn\kln\rightarrow\pi^+\pi^- \, \pi^{\pm} e^{\mp} \nu $ and 
$\phi\rightarrow\ksn\kln\rightarrow \pi^{\pm} e^{\mp} \nu \, 3\pi^0$ processes,
%decay channels and their $\Delta t$ distributions 
with $\Delta t$ the difference of the kaon decay times.
A comparison of the measured $\Delta t$ distributions in the asymptotic region $\Delta t \gg \tau_S$
allows to test for the first time \T and \CPT symmetries in kaon transitions with a precision of few percent,
and to observe \CP violation with this novel method.
\end{abstract}




\end{frontmatter}

\input introduction
\input experimental

\input conclusions
% 

\section*{Acknowledgments}
We warmly thank our former KLOE colleagues for the access to the data collected during the KLOE data taking campaign.
We thank the DA$\Phi$NE team for their efforts in maintaining good running conditions
and their collaboration during both the KLOE run %\cite{DAPHNE1} 
and the KLOE-2 data taking with an upgraded collision scheme \cite{DAPHNE2, DAPHNE3}.
We are very grateful to our colleague G. Capon for his enlightening comments and suggestions about the manuscript.
We want to thank our technical staff: 
G.F. Fortugno and F. Sborzacchi for their dedication in ensuring efficient operation of the KLOE computing facilities; 
M. Anelli for his continuous attention to the gas system and detector safety; 
A. Balla, M. Gatta, G. Corradi and G. Papalino for electronics maintenance; 
C. Piscitelli for his help during major maintenance periods. 
This work was supported in part 
by the Polish National Science Centre through the Grants No.
2014/14/E/ST2/00262,
2016/21/N/ST2/01727,
2017/26/M/ST2/00697.
A.G. acknowledges the support from the Foundation for Polish Science through grant TEAM POIR.04.04.00-00-4204/17.

% \section{}
%\label{}




\bibliographystyle{elsarticle-num} 
\bibliography{refs.bib, intro_refs.bib}

%\input biblio_5
%
%\begin{thebibliography}{00}


%\bibitem{}

%\end{thebibliography}
\end{document}

%% file: authors.tex
\author{The KLOE-2 Collaboration}
\author[Frascati]{D.~Babusci}
\author[Warsaw]{M.~Ber{\l{}}owski}
\author[Frascati]{C.~Bloise}
\author[Frascati]{F.~Bossi}
\author[INFNRoma3]{P.~Branchini}
\author[Uppsala]{B.~Cao}
\author[Roma3,INFNRoma3]{F.~Ceradini}
\author[Frascati]{P.~Ciambrone}
\author[Calabria,INFNCalabria]{F.~Curciarello}
\author[Cracow]{E.~Czerwi\'nski}
\author[Roma1,INFNRoma1]{G.~D'Agostini}
\author[Roma1,INFNRoma1]{R.~D'Amico}
\author[Frascati]{E.~Dan\`e}
\author[Roma1,INFNRoma1]{V.~De~Leo}
\author[Frascati]{E.~De~Lucia}
\author[Frascati]{A.~De~Santis}
\author[Frascati]{P.~De~Simone}
\author[Roma1,INFNRoma1]{A.~Di~Domenico}
\author[Frascati]{E.~Diociaiuti}
\author[Frascati]{D.~Domenici}
\author[Frascati]{A.~D'Uffizi}
\author[Roma1,INFNRoma1]{G.~Fantini}
\author[Cracow]{A.~Gajos}
\author[Cracow]{S.~Gamrat}
\author[Roma1,INFNRoma1]{P.~Gauzzi}
\author[Frascati]{S.~Giovannella}
\author[INFNRoma3]{E.~Graziani}
\author[Wuhan]{X.~Kang}                        % removed LNF on 11-Jul-2022
\author[Uppsala,Warsaw]{A.~Kup\'s\'c}              % added Warsaw on 11-Jul-2022
\author[Messina,INFNCatania]{G.~Mandaglio}
\author[Frascati,Marconi]{M.~Martini}
\author[Frascati]{S.~Miscetti}
\author[Cracow]{P.~Moskal}
\author[INFNRoma3]{A.~Passeri}
\author[Cracow]{E.~P\'erez~del~R\'{\i}o}
\author[Calabria,INFNCalabria]{M.~Schioppa}
\author[Roma3,INFNRoma3]{A.~Selce}
\author[Cracow]{M.~Silarski}
\author[Frascati,IFIN]{F.~Sirghi}
\author[BINP,Novosibirsk]{E.~P.~Solodov}
\author[Warsaw]{W.~Wi\'slicki}
\author[Uppsala]{M.~Wolke}
%\author{\\and\\} 
\author[Valencia]{\\and J.~Bernab{\'e}u}
%%%%%%%%%%%%%%%%%%%%%%%%%%%%%%%%%%%%%%%%%%%%%%%%%%%%%%%%%%%%%%%%%%%%%%%%%%%%%%%%%%%%%%%%%%%%%%%%%%
\address[INFNCatania]{\affinfn{Catania}{Catania}}
\address[Cracow]{Institute of Physics, Jagiellonian University, Cracow, Poland.}
\address[Frascati]{Laboratori Nazionali di Frascati dell'INFN, Frascati, Italy.}
\address[IFIN]{Horia Hulubei National Institute of Physics and Nuclear Engineering, M\v{a}gurele, Romania.}
\address[Messina]{Dipartimento di Scienze Matematiche e Informatiche, Scienze Fisiche e Scienze della Terra dell'Universit\`a di Messina, Messina, Italy.}
\address[BINP]{Budker Institute of Nuclear Physics, Novosibirsk, Russia.}
\address[Novosibirsk]{Novosibirsk State University, Novosibirsk, Russia.}
\address[Calabria]{\affuni{della Calabria}{Arcavacata di Rende}}
\address[INFNCalabria]{INFN Gruppo collegato di Cosenza, Arcavacata di Rende, Italy.}
\address[Marconi]{Dipartimento di Scienze e Tecnologie applicate, Universit\`a ``Guglielmo Marconi'', Roma, Italy.}
\address[Roma1]{\affuni{``Sapienza''}{Roma}}
\address[INFNRoma1]{\affinfn{Roma}{Roma}}
\address[Roma3]{Dipartimento di Matematica e Fisica dell'Universit\`a ``Roma Tre'', Roma, Italy.}
\address[INFNRoma3]{\affinfn{Roma Tre}{Roma}}
\address[Uppsala]{Department of Physics and Astronomy, Uppsala University, Uppsala, Sweden.}
\address[Valencia]{Department of Theoretical Physics, University of Valencia, and IFIC, Univ. Valencia-CSIC, E-46100 Burjassot, Valencia, Spain.}
\address[Wuhan]{School of Mathematics and Physics, China University of Geosciences (Wuhan), Wuhan, China.}
\address[Warsaw]{National Centre for Nuclear Research, Warsaw, Poland.}

%% file: introduction.tex
\section{Introduction \label{sec:introduction}}
Symmetries and their breaking in the physical laws play a crucial role in fundamental physics
and other fields. Besides local gauge symmetries generated by charges 
the discrete symmetries are also of maximal relevance. The breaking of Parity -- \P, Charge Conjugation -- \C, and the combined \CP symmetries 
%through definite
%patterns 
has oriented the understanding of the electroweak and flavour sectors of particle
physics. Whereas \P, \C and \CP are implemented by unitary operators, 
Time Reversal \T and \CPT transformations are antiunitary~\cite{wigner}. 
This fact implies  that, besides the transformation  of
initial and final states in a given process, the two states have to be exchanged for a genuine
direct test of the symmetry. For unstable particles this requirement, being the decay
irreversible for all practical purposes, seems to lead to a \textit{no-go} argument~\cite{wolf,wolf2}. 
%This 
The latter can be bypassed
considering that the {\it reversal-in-time} does not include the decay products,
but only the {\it motion-reversal} before the decay, the decay being instrumental for tagging the
initial meson state and filtering the final state~\cite{banuls1,banuls2}. This is 
%In particular the tagging of the state is
%In fact tagging and filtering of the states are made possible
%, and 
made possible
%In fact 
by 
exploiting the maximal entanglement of 
meson--antimeson pairs, as 
$\bn\bnb$ 
%pairs
produced at \bbn-factories, or $\kn\knb$ 
%pairs 
produced at $\phi$-factories.
%\T violation (TRV) effects
This conceptual basis for the search of
Time Reversal Violation (TRV) effects
was immediately recognised~\cite{wolf} as being a genuine test. 
%As a
%consequence, the experiment has to control the connection of the measured observables to the
%{\it theoretical} asymmetry for transition probabilities  between meson states.
%
%,
% the role of the decay can be only
% %being 
% instrumental for tagging the
%initial meson state and filter the final one~\cite{banuls1,banuls2}. 
The methodology for the entire procedure was developed in Ref.~\cite{btviol} for the \bn-system,
yielding the first direct observation of TRV by the BABAR Collaboration \cite{babartviol} 
(additional information can be found in~Refs.~\cite{applebaum,genuinet}).
%for a detailed analysis)
For the K-system, the methodology for the test
in transitions involving 
meson decays into specific flavour or \CP final states
was described in Refs.~\cite{ktviol,kcptviol}.  
%specific decay products of flavour and \CP eigenstates.
% decay products.
%\par
%The first direct observation of TRV following these concepts and method
%%has been 
%was made in the
%$\bn-\bnb$ system by the BABAR Collaboration \cite{babartviol}.
%A  detailed analysis of the measured time-dependent
%asymmetries \cite{applebaum,genuinet} in the framework of the Weisskopf-Wigner effective
%Hamiltonian approach \cite{ww} for the $\bn-\bnb$ system confirmed the
%TRV effect and identified the precise information contained in
%the asymmetry parameters of the different time-dependent terms.
%% in the time-dependence.
\par
    For the 
    %neutral K-meson
    $\kn-\knb$ system, the CPLEAR Collaboration obtained a 
%    four sigma 
$4\sigma$    
    evidence of
%    measured
the 
%interesting 
asymmetry between the 
%CP-conjugate 
%\CPT even
process
%transition 
$\kn\rightarrow\knb$ and its inverse $\knb\rightarrow\kn$~\cite{cpleartviol}.
%, therefore identical for \T and \CP tests. 
However, its interpretation in terms of a genuine TRV effect is controversial,
%problematic, 
raising some issues \cite{wolf,wolf2,bernabeucolloquium}
%[3, 11,12] 
related to the decay as essential
ingredient. 
%The effect needs 
% for the test
%of the decay 
%leading to 
In fact 
an initial state absorptive interaction between
\kn 
%K0 
and 
\knb
%K0bar 
is needed to generate the TRV effect in this case,
%to generate the asymmetry, that is
leading to
%and 
an asymmetry 
constant in time,
due to
the non-orthogonality of $\kln$ and $\ksn$. 
Moreover, for the 
$\kn\rightarrow\knb$
transition its \CPT transformed is the identity, 
not distinguishing \T and \CP conjugate transitions, even if \CPT is violated.
% the specific process observed, that is \CPT even, makes experimentally undistinguishable 
%\T from \CP violation effects. 
The related \CPT test performed by CPLEAR is based on the comparison of 
the $\kn\rightarrow\kn$ and $\knb\rightarrow\knb$ processes, i.e. %of 
the
\kn and \knb
survival probabilities~\cite{cplearcptviol}.
\par
In this paper we present the results of the first direct 
%genuine 
\T, \CP, \CPT 
tests performed in the $\kn-\knb$ system, 
obtained analysing the data collected by the KLOE experiment at the 
DA$\Phi$NE $\phi$-factory, and
according to the methodology described in Refs.\cite{ktviol,kcptviol}. 
The transitions involved are those
between the 
%\knn and \kn, \knb 
states tagged and filtered by the \CP and
flavour eigenstate decay products. In different combinations, they
allow to build direct, genuine and separate observables for \T, \CP
and \CPT asymmetries.
In particular, the test
%\CPT test  
based on the measurement of the double ratio 
$DR_{\CPT}$
(Eq.~(\ref{eq:ratiosdef8}))
%\ref{doubleratiodslimit}))
%in the neutral kaon system 
is 
%one of the cleanest ever performed.
very clean, free from approximations and model independent, and constitutes an excellent tool
 to advance in the search
for \CPT violation.
% a breaking of the \CPT
%this 
%symmetry. 
In fact \CPT invariance has very solid theoretical justification in
the \CPT theorem 
\cite{luders,pauli,bellcpt,jost},
valid for
quantum field theories satisfying Lorentz invariance, local interaction and unitarity,
and an unambiguous observation of its violation would have grave consequences 
for our understanding of particle physics.
Probing \CPT in transitions selects a different sensitivity to violating effects with respect to the test~\cite{cplearcptviol} based on survival probabilities 
for \kn and \knb  involving diagonal terms
of the Hamiltonian.
\section{Transition probabilities and double decay intensities}
\par
In order 
to implement 
the 
%a direct test of the 
\T, \CP and \CPT tests,
%symmetries,
%%\T and 
%\CPT symmetry,
%, which
%requires the inversion of {\it in} and {\it out} states as in the test of \T symmetry \cite{tviol},
%it has been suggested to exploit 
the 
%Einstein-Podolsky-Rosen (EPR) 
%\cite{ref:EPR}
entanglement of neutral kaons
%\footnote{A fascinating subject still recently investigated both experimentally by the KLOE-2 collaboration~\cite{kloeqm2} and theoretically~\cite{future}.}
produced at DA$\Phi$NE (see recent experimental \cite{kloeqm2} and theoretical studies~\cite{future} on this subject)
%by the KLOE-2 collaboration~
%\cite{kloeqm2},
%a $\phi$-factory 
is exploited.
%(or B-factory\footnote{It's worth mentioning that
%in the neutral B meson system a direct \T test has been already accomplished \cite{ref:babarTviol}, exploiting a similar methodology as discussed in the following.}) \cite{ref:bernabeuPLB-NPB,ref:BernabeuDiscrete,ref:Bmethod,tviol}.
In fact in this case
the initial state of the neutral kaon pair produced in $\phi\rightarrow \kn\knb$ decay
can be rewritten in terms of any pair of orthogonal states
$|\kpp \rangle$ and $|\knn \rangle$ as:
\begin{eqnarray}
  |i \rangle   =  \frac{1}{\sqrt{2}} \{ |\kn \rangle |\knb \rangle - 
 |\knb \rangle |\kn \rangle
\} 
\label{eq:state1}
   =  \frac{1}{\sqrt{2}} \{ |\kpp \rangle |\knn \rangle - 
 |\knn \rangle |\kpp \rangle
\label{eq:state3}
\}~.
\end{eqnarray}  
%Here one can 
%%%%%%%%%%%%%%%%
\par
In order to formulate the test, it is necessary to precisely define the different states involved 
as {\it in} and {\it out} states in the considered transition process.
%as possible 
%in and out states in the time evolution.
%In order to implement the test, two orthogonal bases 
%of the kaon states involved in the
%transition process have to be defined .
%is first needed to define the 
%two orthogonal bases subjected to
%whose states to be 
%consider as the
%
First, let us consider the states $|\knn\rangle$, $|\kpp\rangle$ defined as follows:
$|\knn\rangle$, $|\kpp\rangle$ are the states tagged 
%for the partner 
by the observation of the partner decay into 
%of 
the $\CP=+, -$ eigenstate decay products $\pi^+\pi^-$ 
%($\pi^+\pi^-$ 
(or equivalently $\pi^0\pi^0$) and $3\pi^0$, respectively. 
They are explicitly identified as the states not
decaying to these channels:
%
%the following channels, respectively:
%$|\kpp\rangle$ is the state filtered by the decay into $\pi^+\pi^-$ 
%%($\pi^+\pi^-$ 
%(or equivalently $\pi^0\pi^0$), a
%pure $\CP=+1$ state; analogously $|\knn\rangle$ is the state filtered by the decay into 
%$3\pi^0$, a
%pure $\CP=-1$ state. 
%Their orthogonal states correspond to the states which cannot decay into 
%$\pi\pi$ or $3\pi^0$, defined, respectively, as
\begin{eqnarray}
|\knn\rangle &\propto&  | \kln\rangle 
- \eta_{+-}|\ksn \rangle  \nonumber \\
|\kpp\rangle &\propto& | \ksn\rangle 
- \eta_{000} |\kln \rangle ~, 
\end{eqnarray}
with
$\eta_{+-}={\langle \pi^+\pi^- |T |\kln\rangle}/{\langle \pi^+\pi^- |T |\ksn\rangle}$
and 
%$\eta_{3\pi^0}={\langle 3\pi^0 |T |\ksn\rangle}/{\langle 3\pi^0 |T |\kln\rangle}$.
$\eta_{000}={\langle 3\pi^0 |T |\ksn\rangle}/{\langle 3\pi^0 |T |\kln\rangle}$.
Thus their orthogonal
states 
$|\knnort\rangle$, 
$|\kpport\rangle$
%
%$K-T>$, $/K+T>$ 
are those filtered by their observed decays.
The orthogonality condition  
$|\knnpport\rangle\equiv| \kppnn \rangle$
%/K-+T> = /K+-> 
%
%corresponds to assume
%
%With these definitions of states, it can be shown that 
%the condition of orthogonality $\langle\knn|\kpp\rangle=0$, 
%(i.e. $|\kpp\rangle\equiv|\kppp\rangle$ and
%$|\knn\rangle\equiv|\knnp\rangle$)
corresponds to assume negligible direct \CP (and \CPT) violation contributions,
% (i.e. 
%$\epsilon^{\prime}, \epsilon^{\prime}_{000} \ll \epsilon$),
assumption well satisfied for neutral kaons~\cite{kcptviol}.
%(see detailed discussion in Ref. \cite{cpttran}).
%Appendix A 
%of Ref. \cite{tviol}).
As a second orthogonal basis we consider the two flavour eigenstates $|\kn\rangle$ and $|\knb\rangle$.
The validity of the $\Delta S=\Delta Q$ rule is also assumed~\cite{dsdq},
%\footnote{It is important to underline that
%both assumptions of negligible violations of the $\Delta S=\Delta Q$ rule and direct \CP and \CPT
%can be relaxed for a specific \CPT observable, as discussed below.}, 
so that 
%the two flavour orthogonal eigenstates $|\kn\rangle$ and $|\knb\rangle$ 
these states are identified by the charge of the 
lepton
%electron (or equivalently the muon)
in semileptonic decays, i.e. a $|\kn\rangle$ can decay into
$\pi^-e^+\nu$ (or $\pi^-\mu^+\nu$) and not into $\pi^+e^-\bar{\nu}$ (or $\pi^+\mu^-\bar{\nu}$), and vice versa for a $|\knb\rangle$.
%\footnote{in the following the choice $\ell=e$ will be intended, as particle identification techniques work better 
%for electrons than muons at DA KLOE. can be are easier to be detected.}.
\par
Thus, exploiting the perfect anticorrelation of the states implied 
by Eq.~(\ref{eq:state3}),
 it is possible to have a 
\textquotedblleft flavour-tag\textquotedblright or a 
 \textquotedblleft CP-tag\textquotedblright,
% ~\cite{ref:bernabeuJHEP},
i.e.~to infer the flavour (\kn or \knb) or the \CP (\kpp or \knn) state
of the still alive kaon by observing a specific flavour decay ($\pi^-e^+\nu$
or  $\pi^+e^-\bar{\nu}$) or CP decay ($\pi^+\pi^-$ or $\pi^0\pi^0\pi^0$) of the other (and first decaying) kaon in the pair.

\par
In this way 
one can experimentally access -- for instance -- the transition $\kn \to \kpp$, taken as reference, and
%the 
$\kpp \to \kn$, $\knb \to \kpp$ and $\kpp \to \knb$,
%transitions, 
i.e. the \T, \CP and \CPT
conjugated transitions,
% of the reference one, 
respectively.
%
%listed 
%
All possible transitions 
can be divided
into four categories of events, corresponding to independent \T, \CP and \CPT tests.
%, as listed
%in Table~\ref{tab:processes}.
%\begin{table}[h]
%  \begin{center}
%    \caption{Scheme of possible reference transitions and their associated
%      \T, \CP or \CPT conjugated processes accessible at a $\phi$-factory.
%    \label{tab:processes}}
%    \begin{tabular}{c|c|c|c}		
%      \hline
%%      \multicolumn{2}{c|}{Reference}  &   %\multicolumn{2}{c}{\T-conjugate} \\
%      Reference & T-conjug.   &  CP-conjug. & CPT-conjug. \\ \hline
%      \trule $\kn \to \kpp$   & $\kpp \to \kn$    & $\knb \to \kpp$ %& $\kpp \to \knb$ \\
%      \trule $\kn \to \knn$ &  $\knn \to \kn$ &  $\knb \to \knn$ & % $\knn \to \knb$ \\
%      \trule $\kpp \to \knb$   & $\knb \to \kpp$    &  $\kpp \to %\kn$ & $\kn \to \kpp$ \\
%      \trule $\knn \to \knb$  & $\knb \to \knn$   & $\knn \to \kn$ % & $\kn \to \knn$   \\ \hline
%   \end{tabular}
%  \end{center}	
%\end{table}
%\section{\T and \CPT symmetry test}
%Specifically for the \CPT symmetry test, 
One can directly compare the probabilities for the reference transition
and the conjugated transition defining 
%through
% the ratios of probabilities defined as follows:
%For the direct T symmetry test one can define 
the following ratios of probabilities for the \T 
%symmetry 
test:
\begin{eqnarray}
R_{1,\T}(\Delta t) &=& P\left[\kpp(0)\to\knb(\Delta t)\right]/P\left[\knb(0)\to\kpp(\Delta t)\right] \nonumber \\
R_{2,\T}(\Delta t) &=& P\left[\kn(0)\to\knn(\Delta t)\right]/P\left[\knn(0)\to\kn(\Delta t)\right] \nonumber\\
R_{3,\T}(\Delta t) &=& P\left[\kpp(0)\to\kn(\Delta t)\right]/P\left[\kn(0)\to\kpp(\Delta t)\right] \nonumber\\
R_{4,\T}(\Delta t) &=& P\left[\knb(0)\to\knn(\Delta t)\right]/P\left[\knn(0)\to\knb(\Delta t)\right]~,
\label{eq:tratios}
\end{eqnarray}
for the \CP 
%symmetry 
test:
\begin{eqnarray}
R_{1,\CP}(\Delta t) &=& P\left[\kpp(0)\to\knb(\Delta t)\right]/P\left[\kpp(0)\to\kn(\Delta t)\right] \nonumber \\
R_{2,\CP}(\Delta t) &=& P\left[\kn(0)\to\knn(\Delta t)\right]/P\left[\knb(0)\to\knn(\Delta t)\right] \nonumber\\
R_{3,\CP}(\Delta t) &=& P\left[\knb(0)\to\kpp(\Delta t)\right]/P\left[\kn(0)\to\kpp(\Delta t)\right] \nonumber\\
R_{4,\CP}(\Delta t) &=& P\left[\knn(0)\to\kn(\Delta t)\right]/P\left[\knn(0)\to\knb(\Delta t)\right]~,
\label{eq:cpratios}
\end{eqnarray}
or for the \CPT 
%symmetry 
test:
\begin{eqnarray}
R_{1,\CPT}(\Delta t) &=& P\left[\kpp(0)\to\knb(\Delta t)\right]/P\left[\kn(0)\to\kpp(\Delta t)\right] \nonumber \\
R_{2,\CPT}(\Delta t) &=& P\left[\kn(0)\to\knn(\Delta t)\right]/P\left[\knn(0)\to\knb(\Delta t)\right] \nonumber\\
R_{3,\CPT}(\Delta t) &=& P\left[\kpp(0)\to\kn(\Delta t)\right]/P\left[\knb(0)\to\kpp(\Delta t)\right] \nonumber\\
R_{4,\CPT}(\Delta t) &=& P\left[\knb(0)\to\knn(\Delta t)\right]/P\left[\knn(0)\to\kn(\Delta t)\right]~.
\label{eq:cptratios}
\end{eqnarray}

The measurement of any deviation from the prediction $R_{i,\SS}(\Delta t)=1$
%with $\SS=\T,\CP,\CPT$, 
%\begin{eqnarray}
%R_1(\Delta t)=R_2(\Delta t)=R_3(\Delta t)=R_4(\Delta t)=1
%\label{eq:tprediction}
%\end{eqnarray}
imposed by the symmetry invariance (with $\SS=\T,\ \CP$ or $\CPT$ and $i=1,4$)
%\par The measurement of any $R_i\neq 1$ 
is a direct and genuine signal of the corresponding symmetry
violation.
%is a clean and direct signal of the symmetry violation. 
\\
It is worth noting that for $\Delta t=0$:
\begin{linenomath*}
\begin{eqnarray}
R_{1,\SS}(0)=R_{2,\SS}(0)=R_{3,\SS}(0)=R_{4,\SS}(0)=1~,
\label{eq:one}
\end{eqnarray}
\end{linenomath*}
i.e. the 
%\SS-
\T, \CP, \CPT
violating effects are built in the time 
evolution of the system, and 
are
%it is 
absent at $\Delta t=0$, within our approximations.
For $\Delta t\gg \tau_S$ all ratios,
including the symmetry violating effects,
reach an asymptotic regime.
\par
At a $\phi$-factory 
one can define two observable ratios for each symmetry test~\cite{ktviol,kcptviol,didoactapp}:
%
%the corresponding observable quantities are two ratios,
%$R_{2,\CPT}^{\rm{exp}}(\Delta t)$ and $R_{4,\CPT}^{\rm{exp}}(\Delta t)$,
%For $\Delta t >0$ one has:
%
%one can define the observable ratios:
\begin{eqnarray}
\label{ratio2t}
R_{2,\T}^{\rm{exp}}(\Delta t) \equiv
\frac{  I(\pi^+e^-\bar{\nu},3\pi^0;\Delta t)}
{ I(\pi^+\pi^-,\pi^-e^+\nu;\Delta t)}  
%\end{eqnarray} 
%\end{linenomath*}
%\begin{linenomath*}
%\begin{eqnarray}
\label{ratio4t}
~;~
R_{4,\T}^{\rm{exp}}(\Delta t) \equiv
\frac{  I(\pi^-e^+\nu,3\pi^0;\Delta t)}
{ I(\pi^+\pi^-,\pi^+e^-\bar{\nu};\Delta t)}   
%~,
%=\frac{R_{4,\CPT}(\Delta t)}{D_{\CPT}}
%= R_2(\Delta t)
%&\prop&
%P\left[ \knb(0)\to\knn(t_2-t_1) \right]
\end{eqnarray} 
%\end{linenomath*}
\begin{linenomath*}
\begin{eqnarray}
\label{ratio2cp}
R_{2,\CP}^{\rm{exp}}(\Delta t) \equiv
\frac{  I(\pi^+e^-\bar{\nu},3\pi^0;\Delta t)}
{ I(\pi^-e^+\nu,3\pi^0;\Delta t)}
%\end{eqnarray} 
%\end{linenomath*}
%\begin{linenomath*}
%\begin{eqnarray}
\label{ratio4cp}
~;~
R_{4,\CP}^{\rm{exp}}(\Delta t) \equiv
\frac{  I(\pi^+\pi^-,\pi^-e^+\nu;\Delta t)}
{ I(\pi^+\pi^-,\pi^+e^-\bar{\nu};\Delta t)}   
%~,
%=\frac{R_{4,\CPT}(\Delta t)}{D_{\CPT}}
%= R_2(\Delta t)
%&\prop&
%P\left[ \knb(0)\to\knn(t_2-t_1) \right]
\end{eqnarray} 
\end{linenomath*}
\begin{linenomath*}
\begin{eqnarray}
\label{ratio2cpt}
R_{2,\CPT}^{\rm{exp}}(\Delta t) \equiv
\frac{  I(\pi^+e^-\bar{\nu},3\pi^0;\Delta t)}
{ I(\pi^+\pi^-,\pi^+e^-\bar{\nu};\Delta t)}
%\end{eqnarray} 
%\end{linenomath*}
%\begin{linenomath*}
%\begin{eqnarray}
\label{ratio4cpt}
~;~
R_{4,\CPT}^{\rm{exp}}(\Delta t) \equiv
\frac{  I(\pi^-e^+\nu,3\pi^0;\Delta t)}
{ I(\pi^+\pi^-,\pi^-e^+\nu;\Delta t)}   ~,
%=\frac{R_{4,\CPT}(\Delta t)}{D_{\CPT}}
%= R_2(\Delta t)
%&\prop&
%P\left[ \knb(0)\to\knn(t_2-t_1) \right]
\end{eqnarray} 
\end{linenomath*}
where
$I(f_1,f_2;\Delta t)$
is the double decay rate
into 
%specific 
%decay products
final states 
$f_1$ 
%(e.g. for the kaon on the left side) 
and $f_2$ 
%(e.g. for the kaon on the right side) 
as a function of the difference of kaon decay times $\Delta t$ \cite{didohand}, with
$f_1$ occurring before $f_2$ decay for $\Delta t >0$, and vice versa for $\Delta t<0$.
\\
They are related to the $R_{i,\SS}(\Delta t)$ ratios defined in Eqs.~(\ref{eq:tratios}, \ref{eq:cpratios}, \ref{eq:cptratios}) as follows, for $\Delta t  \geq 0$:
%Assuming that $\langle\knn|\kpp\rangle=0$, i.e. $|\kpp\rangle\equiv|\kppp\rangle$ and
%$|\knn\rangle\equiv|\knnp\rangle$, one has
% one has (the first decay product symbol in parenthesis indicates the first of the two kaon decay):
\begin{linenomath*}
\begin{eqnarray}
\label{eq:intensity2}
%\equiv
%\frac{  I(\ell^-,3\pi^0;\Delta t)}
%{ I(\pi\pi,\ell^-;\Delta t)}   
R_{2,\T}^{\rm{exp}}(\Delta t) 
&=&{R_{2,\T}(\Delta t)}\times{D}
\nonumber\\
R_{4,\T}^{\rm{exp}}(\Delta t) 
&=&{R_{4,\T}(\Delta t)}\times{D}
\nonumber\\
R_{2,\CP}^{\rm{exp}}(\Delta t) 
&=&{R_{2,\CP}(\Delta t)}
%\times{D}
\nonumber\\
R_{4,\CP}^{\rm{exp}}(\Delta t) 
&=&{R_{4,\CP}(\Delta t)}
%\times{D}
\nonumber\\
R_{2,\CPT}^{\rm{exp}}(\Delta t) 
&=&{R_{2,\CPT}(\Delta t)}\times{D}
\nonumber\\
R_{4,\CPT}^{\rm{exp}}(\Delta t) 
&=&{R_{4,\CPT}(\Delta t)}\times{D}
\end{eqnarray}  
\end{linenomath*}
%while 
whereas
for $\Delta t <0$ one has:
\begin{linenomath*}
\begin{eqnarray}
\label{eq:intensity2_negative_dt}
R_{2,\T}^{\rm{exp}}(\Delta t) 
&=&{R_{1,\T}(|\Delta t|)}\times{D}
\nonumber\\
R_{4,\T}^{\rm{exp}}(\Delta t) 
&=&{R_{3,\T}(|\Delta t|)}\times{D}
\nonumber\\
R_{2,\CP}^{\rm{exp}}(\Delta t) 
&=&{R_{1,\CP}(|\Delta t|)}
%\times{D}
\nonumber\\
R_{4,\CP}^{\rm{exp}}(\Delta t) 
&=&{R_{3,\CP}(|\Delta t|)}
%\times{D}
\nonumber\\
R_{2,\CPT}^{\rm{exp}}(\Delta t) 
&=&{R_{1,\CPT}(|\Delta t|)}\times{D}
\nonumber\\
R_{4,\CPT}^{\rm{exp}}(\Delta t) 
&=&{R_{3,\CPT}(|\Delta t|)}\times{D}
%R_{2,\SS}^{\rm{exp}}(\Delta t) 
%\equiv
%\frac{  I(\ell^-,3\pi^0;\Delta t)}
%{ I(\pi\pi,\ell^+;\Delta t)}   
%R_{2,\SS}^{\rm{exp}}(\Delta t) 
%&=&{R_{1,\SS}(|\Delta t|)}\times{D_{2,\SS}}
%\nonumber\\
%%~~,~~
%R_{4,\SS}^{\rm{exp}}(\Delta t) 
%%\equiv
%%\frac{  I(\ell^+,3\pi^0;\Delta t)}
%%{ I(\pi\pi,\ell^-;\Delta t)}   
%&=&{R_{3,\SS}(|\Delta t|)}\times{D_{4,\SS}}~,
%%= R_2(\Delta t)
%&\prop&
%P\left[ \knb(0)\to\knn(t_2-t_1) \right]
\end{eqnarray}  
\end{linenomath*}
with $D$ a constant factor given by~\cite{ktviol,kcptviol,didoactapp}:
%$D_{i,\SS}$ constant factors with the following relations among them
%%the constant terms $D_{i,\SS}$
%\cite{didoactapp}:
%%\begin{eqnarray}  
%% D_{\CPT}= \frac{  C(\ell^-,3\pi^0)}
%%{ C(\pi\pi,\ell^-)}  = \frac{  C(\ell^+,3\pi^0)}
%%{ C(\pi\pi,\ell^+)} 
%%= \frac{\left| 
%%\langle 3\pi^0|T| \knn \rangle\right|^2}{ \left|\langle \pi\pi|T| \kpp \rangle\right|^2}
%%\end{eqnarray}  
%\begin{eqnarray}
%D_{2,\T}=\left( 1+4\Re y \right) \times D_{\CPT}
%&~;~&
%%\nonumber \\
%D_{4,\T}= \left( 1-4\Re y \right) \times D_{\CPT}
%\nonumber \\
%D_{2,\CP}=\left( 1+4\Re y \right)
%%\nonumber \\
%&~;~&
%D_{4,\CP}=\left( 1-4\Re y \right)
%\nonumber \\
%D_{2,\CPT}= D_{\CPT}
%%\nonumber \\
%&~;~&
%D_{4,\CPT}= D_{\CPT}
%\end{eqnarray}
%%with 
%%$D_{\CPT}$ 
%%the ratio of coefficients:
%with the small parameter $y$ describing a possible \CPT violation in the $\Delta S = \Delta Q$ semileptonic decay amplitudes,
%and
\begin{linenomath*}
\begin{eqnarray}  
 D
 %_{\CPT}
 %\frac{  C(\ell^-,3\pi^0;\Delta t)}
%{ C(\pi\pi,\ell^-;\Delta t)}  = \frac{  C(\ell^+,3\pi^0;\Delta t)}
%{ C(\pi\pi,\ell^+;\Delta t)} 
= \frac{\left| 
\langle 3\pi^0|T| \knn \rangle\right|^2}{ \left|\langle \pi^+\pi^-|T| \kpp \rangle\right|^2} 
=\frac{ {\rm BR} \left( \kln\rightarrow 3\pi^0\right) }{ {\rm BR}\left( \ksn\rightarrow \pi^+\pi^-\right)}
\frac{\Gamma_L}{\Gamma_S}~.
\end{eqnarray}  
\end{linenomath*}
%where 
The last r.h.s. equality holds with a high degree of accuracy, at least $\mathcal{O}(10^{-7})$.
The value of 
$D$
%$D_{\CPT}$ 
can be
therefore directly evaluated 
%directly measurable 
from branching ratios and lifetimes of \ksln states.
%%%%%%%%%%%%%%%%%%%
%As shown in Equations~\ref{eq:r2t} and~\ref{eq:r4t}, the experimentally-observable ratios of double kaon decay rates are linked to the theoretical probability ratios with a proportionality constant $D$. With the assumptions discussed in Reference~\cite{theory:bernabeu-t}, the D factor can be expressed completely in terms of experimentally-measured properties of the neutral kaon system:
%\begin{equation}
%  \label{eq:D}
%  D  = \frac{\text{BR}(\Kl\to 3\pi^0)\tau_S}{\text{BR}(\Ks\to\pi^+\pi^-)\tau_L}.
%\end{equation}
%
%These four quantities
% four values entering its calculation 
They
were all directly measured by the KLOE experiment with the highest precision \cite{kllifetime,kl3pi0,kspp,kslifetime,pdg}, 
%with the exception of the \ksn lifetime~\cite{kslifetime}, that anyhow has a very good precision giving a negligible contribution to the 
%uncertainty,
and we consistently use 
them 
for the evaluation 
of $D=(0.5076\pm0.0059)\times 10^{-3}$. 
%_{\CPT}
\par
A Monte Carlo simulation shows that in the case of KLOE and KLOE-2 experiments with an integrated luminosity of $\mathcal{O}(10~{\rm fb}^{-1})$ 
the statistically most populated region is for $\Delta t\gg \tau_S$, while the region for $\Delta t < 0$ has few or no events~\cite{ktviol}, due to unfavourable \ksln branching ratios especially when involving the transitions $\kpp \rightarrow \kn, \knb$. 

%All ratios 
Therefore we define eight observables (six ratios and two double ratios) that are experimentally 
accessible at KLOE with positive $\Delta t$ in the asymptotic regime\footnote{See Ref.~\cite{addcpt2022} for more general formulae valid for $\Delta t\geq0$.}:
% $\Delta t\gg \tau_S$:
\begin{linenomath*}
\begin{eqnarray}
%\equiv
%\frac{  I(\ell^-,3\pi^0;\Delta t)}
%{ I(\pi\pi,\ell^-;\Delta t)}   
\label{eq:ratiosdef1}
R_{2,\T} & \equiv & \frac{R_{2,\T}^{\rm{exp}}(\Delta t\gg \tau_S)}{D} = 1-4\Re \epsilon + \big[4\Re x_+ + 4 \Re y \big]
\\
\label{eq:ratiosdef2}
R_{4,\T} & \equiv & \frac{R_{4,\T}^{\rm{exp}}(\Delta t\gg \tau_S)}{D} = 1+4\Re \epsilon + \big[ 4\Re x_+ - 4 \Re y \big]
\\
\label{eq:ratiosdef3}
R_{2,\CP} & \equiv & R_{2,\CP}^{\rm{exp}}(\Delta t\gg \tau_S) = 1-4\Re \epsilon_S  + \big[ 4 \Re y -4\Re x_-  \big]
\\
\label{eq:ratiosdef4}
R_{4,\CP} & \equiv & R_{4,\CP}^{\rm{exp}}(\Delta t\gg \tau_S) = 1+4\Re \epsilon_L -  \big[ 4 \Re y + 4\Re x_-  \big] 
\\
\label{eq:ratiosdef5}
R_{2,\CPT} & \equiv & \frac{R_{2,\CPT}^{\rm{exp}}(\Delta t\gg \tau_S)}{D} = 1-4\Re \delta + \big[ 4\Re x_+ -  4 \Re x_-\big]
\\
\label{eq:ratiosdef6}
R_{4,\CPT} & \equiv & \frac{R_{4,\CPT}^{\rm{exp}}(\Delta t\gg \tau_S)}{D} = 1+4\Re \delta+ \big[ 4\Re x_+ + 4 \Re x_-\big]
\\
\label{eq:ratiosdef7}
DR_{\T,\CP} & \equiv & \frac{R_{2,\T}}{ R_{4,\T}} \equiv \frac{R_{2,\CP}}{R_{4,\CP}} = 1-8\Re\epsilon + \big[ 8 \Re y \big]
\\
\label{eq:ratiosdef8}
DR_{\CPT} & \equiv & \frac{R_{2,\CPT}}{R_{4,\CPT}} = 1-8\Re\delta - \big[ 8 \Re x_- \big]~.
\end{eqnarray}  
\end{linenomath*}
where
%$\epsilon_{S,L}=\epsilon\pm\delta$, 
$\epsilon$ and $\delta$ are the usual \T and \CPT violation parameters in the neutral
kaon mixing, respectively, and $\epsilon_{S,L}=\epsilon\pm\delta$ the \CP impurities in the 
%effective Hamiltonian stationary 
physical states \ksn and \kln;
the small parameter $y$ describes a possible \CPT violation in the $\Delta S = \Delta Q$ semileptonic decay amplitudes, while
$x_+$ and $x_-$ describe $\Delta S \neq \Delta Q$ semileptonic decay amplitudes
%a possible $\Delta S=\Delta Q$ violation in semileptonic decay amplitudes 
with \CPT invariance and \CPT violation, respectively.
\par
The r.h.s. in Eqs.~(\ref{eq:ratiosdef1}-\ref{eq:ratiosdef8})
are therefore
assuming the presence of 
symmetry 
violations only in the effective Hamiltonian of
the Weisskopf--Wigner approach~\cite{ww} 
%and no
%violations only in the mixing parameters in the Weisskopf-Wigner approximation \cite{ww}
%%mass matrix
and no other possible symmetry violation effect.
The square brackets show the small possible spurious effects (to first order in small parameters) due to the release of the  assumptions of the $\Delta S=\Delta Q$ rule and negligible direct \CP and \CPT violation effects.
\par
%While for \T and \CP tests
In case of the \T and \CPT tests attention must be paid to 
the presence of direct \CP violation contributions in the decay amplitudes.
%in the difference 
%$(\epsilon^{\prime}_{3\pi^0} -\epsilon^{\prime}_{\pi\pi}  )$ 
Even though in principle they could mimic \T or \CPT violation effects,
they
%can be shown that it
%. These fake effects 
%mostly in the $\Delta t <0$ region,
%while  
turn out to be fully negligible 
% totally irrelevant 
for the 
asymptotic
%plateau
%
 region $\Delta t \gg \tau_S$ (a detailed description can be found in Refs.~\cite{ktviol,kcptviol}).
 The \CPT test with the double ratio 
$ DR_{\CPT}$ is also not affected in the same region by possible 
%The effect of a possible 
violation of the $\Delta S = \Delta Q$ rule,
% is also not affecting the \CPT test 
% in the same region
% with the double ratio 
%$ DR_{\CPT}$, 
because $x_-$ signals also \CPT violation.
% defined as:
%\begin{linenomath*}
%\begin{eqnarray}
%DR_{\CPT}
%%( \Delta t \gg \tau_S) 
%=
%\frac{ R_{2,\CPT}^{\rm{exp}} ( \Delta t \gg \tau_S) }{
%R_{4,\CPT}^{\rm{exp}} ( \Delta t \gg \tau_S) }= 1-8\Re\delta -8 \Re x_- ~,
% \label{doubleratiodslimit}
%\end{eqnarray}
%\end{linenomath*}
%with $x_-$ describing \CPT violation in the $\Delta S \neq \Delta Q$ semileptonic decay amplitudes.
Therefore the double ratio (\ref{eq:ratiosdef8}) 
%that 
constitutes one of the most robust observables of our \CPT test. It is independent of $D$, and
in the limit $\Delta t \gg \tau_S$ it exhibits a pure and genuine \CPT violating effect, even without the assumptions 
of the validity of the $\Delta S = \Delta Q$  rule and of negligible contaminations from direct \CP violation.

%% file: experimental.tex
\section{The KLOE detector}\label{sec:kloe}
The KLOE detector is located
at the DA$\Phi$NE $\phi$-factory~\cite{Gallo:2006yn}, an $e^+e^-$ collider operating
at center-of-mass energy of $m_{\phi}\approx 1.020$~GeV. Collisions predominantly
produce $\phi$ mesons nearly at rest, which decay into an entangled
$\kaon\akaon$ pair in 34\% of cases.

The main components of the detector are a large cylindrical drift chamber~(DC)
and an electromagnetic calorimeter~(EMC),
immersed in the axial magnetic field of 0.52~T
produced by a superconducting coil.
The DC~\cite{Adinolfi:2002uk} uses a low-Z gas mixture of
isobutane~(10\%) and helium~(90\%) assuring transparency
to low-energy photons. % from kaon decays. 
The inner wall of the chamber is made of
light carbon fibre composite and thin (0.1~mm) aluminum foil
to minimise $\Kl\to\Ks$ regeneration.
Vertex reconstruction resolution of the DC is about ~3~mm while
track momentum is reconstructed with $\sigma(p_{T})/p_{T}\approx$~0.4\%.
The large outer radius of the chamber (2~m) allows for recording about 40\%
of all decays of $\Kl$ produced in KLOE.

% UNUSUAL AND REDUNDANT REFERENCE ~\cite{LeeFranzini:2007hj}.

The EMC~\cite{Adinolfi:2002zx} consists of a barrel surrounding
the DC and two endcaps, together covering 98\% of the solid angle.
It contains a total of 2440 cells of 4.4$\times$4.4~cm\textsuperscript{2} cross section,
built of lead foils alternated with scintillating fibres of 1~mm diameter.
Each cell is read out by photomultipliers at both ends.
Energy deposits along the cells are localised
using time difference between photomultiplier signals.
Deposits in adjacent cells are grouped into clusters
for which energy and time resolutions are 
$\sigma(E)/E=0.057/\sqrt{E(\mathrm{GeV})}$ and
$\sigma(t)=54\:\mathrm{ps}/\sqrt{E(\mathrm{GeV})} \oplus 100\:\mathrm{ps}$
and position is resolved with
$\sigma_{\parallel} = 1.4\ {\rm cm}/\sqrt{E\ {\rm (GeV)}}$ along the fibres
and $\sigma_{\perp} = 1.3$ cm in the orthogonal direction.
The acceptance of the EMC is complemented with additional
tile calorimeters covering the two final focusing quadrupole magnets
of the beam line~\cite{Adinolfi:2002me}.

The detector operates with two different triggers~\cite{KLOE:2002mvh};
the calorimeter trigger which requires at least two clusters with E~$>$~50~MeV
in the barrel or E~$>$~150~MeV in the endcaps,
and the DC trigger based on multiplicity and topology of hits in the DC.
%The trigger~\cite{KLOE:2002mvh}
%operates independently on energy deposits in the EMC
%(calorimeter trigger requiring deposits of at least 50~MeV in the EMC barrel and 150~MeV in the endcaps)
%and on multiplicity and topology of hits in the DC cells (DC trigger).
Either of the two triggers is sufficient to start the acquisition of an event.
Subsequently, recorded events are subject to an on-line machine background filter
based on calorimeter information only~\cite{Ambrosino:2004qx}.
Finally, stored events are classified into physics categories based
on a preliminary analysis of their topology.

%%%%%%%%%%%%%%%%%%%%%%%%%%%%%%%%%%%%%%%%%%%%%%%%%%%%%%%%%%%%%%%%%%%%%%%%%%
% Experimental                                                           %
%%%%%%%%%%%%%%%%%%%%%%%%%%%%%%%%%%%%%%%%%%%%%%%%%%%%%%%%%%%%%%%%%%%%%%%%%%
\section{Data analysis}\label{sec:experimental}
The analysis was performed on a dataset collected in the years 2004--2005
with an integrated luminosity of 1.7 fb\textsuperscript{-1}
corresponding to about $1.7 \times 10^{9}$ produced $\Ks\Kl$ pairs
and on a sample of Monte Carlo~(MC) simulated events of the same size.

\subsection{Selection of $\Ks \Kl \to (\pi^{\pm}e^{\mp}\nu)(3\pi^0)$ events}\label{sec:selection1}
$\kaon\to\Km$ and $\akaon\to\Km$ transitions are experimentally identified by events with an early semileptonic
decay of a neutral kaon and a later decay of its entangled partner into 3$\pi^0$.
Presence of a vertex formed by two tracks of opposite charge is required in
a cylindrical volume limited by~$\rho=\sqrt{x^2+y^2} < 3$~cm and $|z|<4.5$~cm
around the average $e^+e^-$ interaction point (IP)
where $z$ is the longitudinal and $x,\ y$ are the transverse coordinates.

The $\Ks\to\pi^+\pi^-$ decays constitute the predominant source of background for semileptonic events.
In order to remove this background we assume that both charged particles are pions,
calculate the invariant mass and reject events with $m_{\pi\pi} > 490$~MeV/c\textsuperscript{2}.

%invariant mass of the decaying particle $m_{\pi\pi}$ was calculated assuming the products are charged pions
%and events with  were rejected.

After reduction of data volume with the preselection
based on $\Ks\to\pi^{\pm}e^{\mp}\nu$ vertex identification, 
the $\Kl\to 3\pi^0\to6\gamma$~decay is reconstructed.
Presence of at least six clusters in the calorimeter not associated to tracks in the DC
each with $\text{E}>20$~MeV is required.
The location and time of the $\Kl$ decay are reconstructed using trilateration~\cite{gajos_thesis}.
While four recorded photons are sufficient for vertex reconstruction,
only events with exclusive detection of all six photons are used in the analysis
to minimise background from $\Ks\to2\pi^0$ which results in a four-photon final state.
Moreover, the requirement of six recorded photons provides additional information
used for enhancing the resolution of vertex reconstruction.

If an event contains seven or more candidate clusters, all 
combinatorial 
choices of six are considered
and 
sets of six photons detected are accepted if the total energy of six clusters
%satisfied $350\:\text{MeV} < \sum_{i=1}^{6}E^{i}_{clu} < 700\: \text{MeV}$
is between 350~MeV and 700~MeV
and the calculated 6$\gamma$ invariant mass is greater than 350~MeV/c\textsuperscript{2}.
Moreover, for each 6-cluster set the decay vertex is reconstructed independently for each choice of 4$\gamma$ subsets
and results are compared providing a test sensitive to presence of photons not originating from the $\Kl$
decay point.
In case of more than one set of six clusters surviving the cuts, the set providing the most consistent
vertex reconstruction 
%amount 
among
its 4$\gamma$ subsets is selected.

\begin{figure}[h!]
  \centering
  \begin{subfigure}{0.4\linewidth}
    {\includegraphics[width=1.0\linewidth]{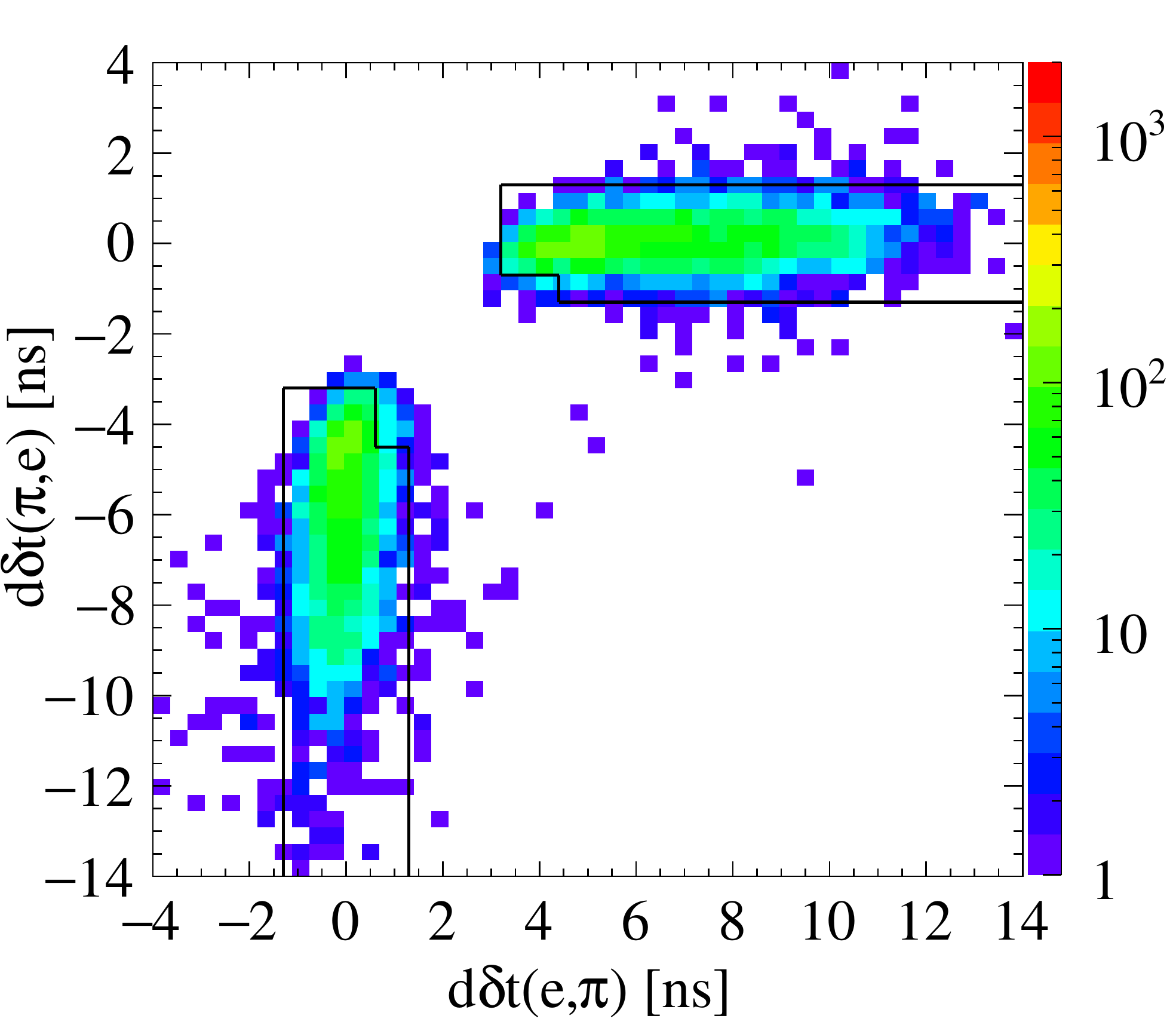}}
  \end{subfigure}
  \begin{subfigure}{0.4\linewidth}
  {\includegraphics[width=1.0\linewidth]{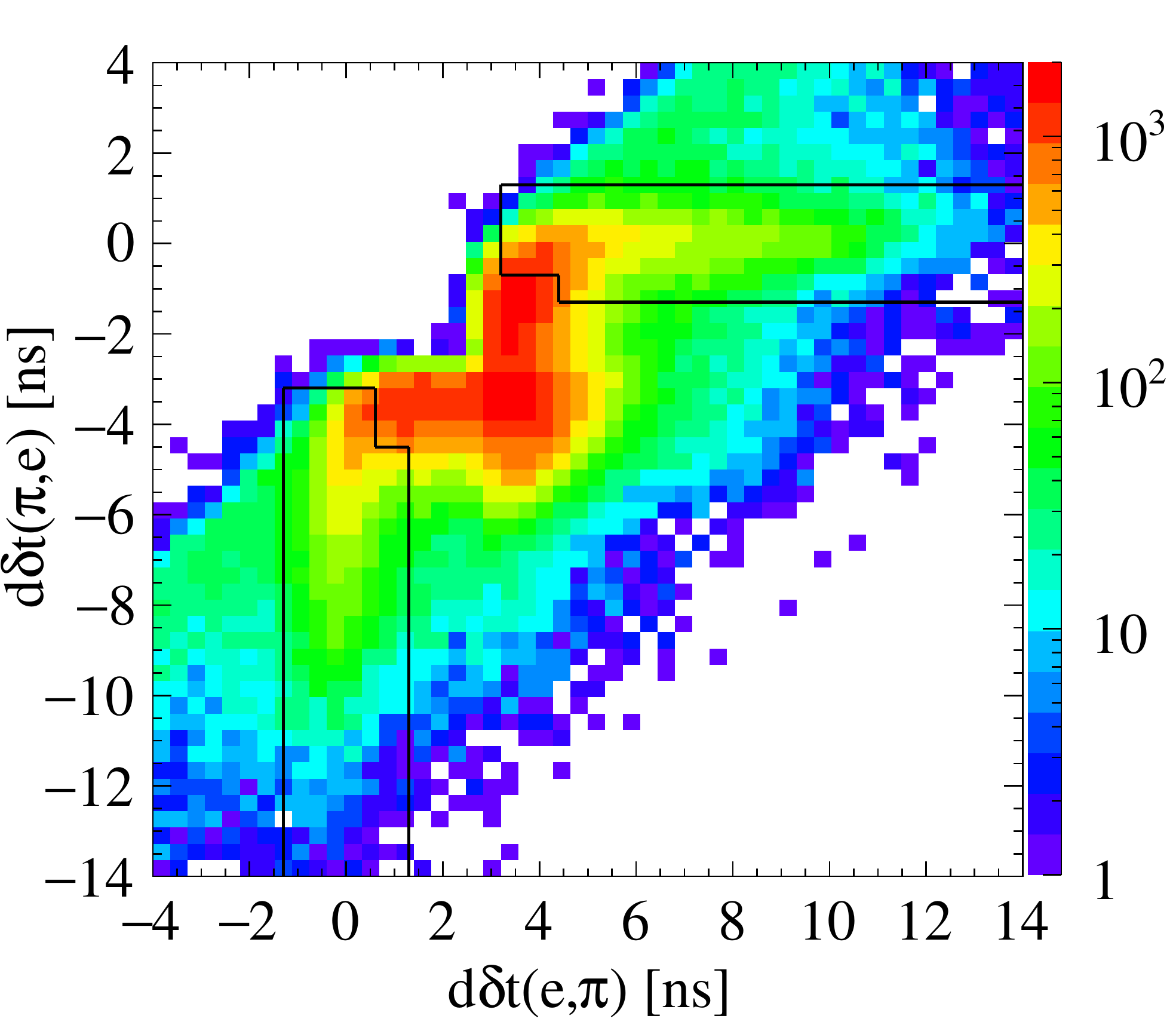}}
\end{subfigure}
\caption{
%Errors of two possible 
Distribution of 
%time difference 
${\rm d}\delta(\pi,e)$ vs ${\rm d}\delta(e,\pi)$ 
%${\rm d}\delta(\pi^{\pm},e^{\mp})$ vs ${\rm d}\delta(e^{\pm},\pi^{\mp})$ 
%$\pi^{\pm}$ and $e^{\mp}$ mass hypotheses assignments to tracks 
for
$\Ks \Kl \to (\pi^{\pm}e^{\mp}\nu)(3\pi^0)$
  MC-simulated 
  %$\Ks\to\pi^{\pm}e^{\mp}\nu$ 
  events (left) and all data events (right).
  Solid lines denote accepted regions, each corresponding to one possible assignment of pion and electron mass hypotheses' to the two DC tracks in an event.}
\label{fig:t1-tof2}
\end{figure}

At this stage, 98\% of selected decays of the second kaon are $3\pi^0$ but the early decays are still dominated by $\pi^+\pi^-$, therefore a time of flight analysis is applied to the tracks of the early decay. The two tracks are extrapolated from the DC to the EMC and differences $\delta t(X)$ between their recorded time of flight $T$ and
time expected from the track length L assuming particle mass $m_X$ are calculated as follows:
\begin{equation}
\label{eq:tof1}
\delta t({m_X}) = T - \frac{L}{c\beta(m_X)}, \quad   \beta(m_X) = \frac{p}{\sqrt{p^{2}+m_X^{2}c^{2}}}.
\end{equation}
Comparison of the differences of $\delta t$ between the two tracks (d$\delta t(X,Y)=\delta t_{trk1}(X) - \delta t_{trk2}(Y)$)
in case of two possible assignments of particle masses shown in Fig.~\ref{fig:t1-tof2}
allows for efficient rejection of $\pi^+\pi^-$ background
as well as for
identification of pion and electron tracks.
%finding the correct attribution of pion and electron masses to the tracks.

\begin{figure}[h!]
  \centering
  \begin{subfigure}{0.4\linewidth}
    {\includegraphics[width=1.0\linewidth]{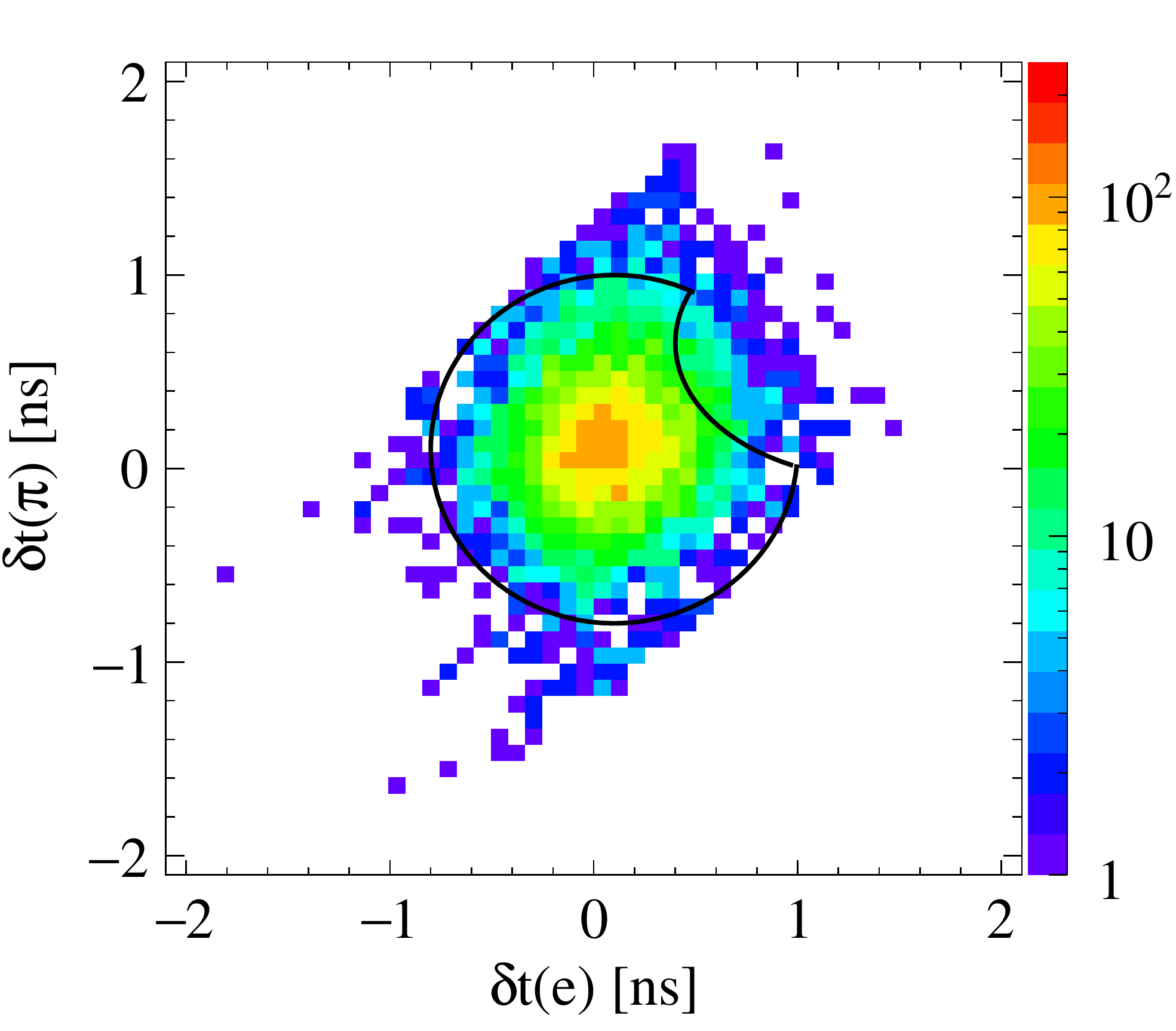}}
  \end{subfigure}
  \begin{subfigure}{0.4\linewidth}
  {\includegraphics[width=1.0\linewidth]{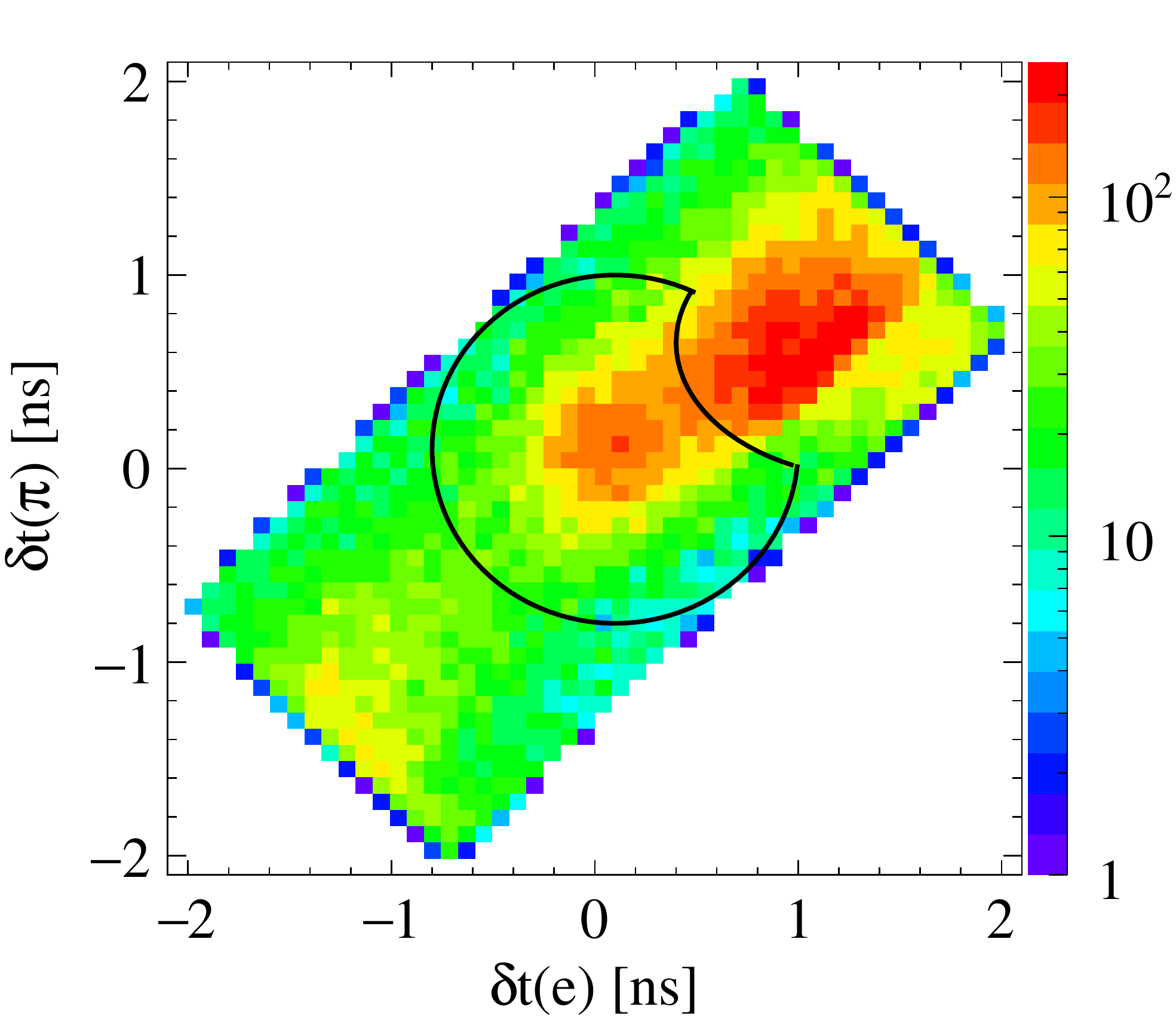}}
\end{subfigure}
\caption{Distribution of 
$\delta t(\pi)$ vs $\delta t(e)$ 
%${\rm d}\delta(\pi^{\pm},e^{\mp})$ vs ${\rm d}\delta(e^{\pm},\pi^{\mp})$ 
%$\pi^{\pm}$ and $e^{\mp}$ mass hypotheses assignments to tracks 
after particle identification
%
%differences between particle time of flight recorded and expected
%  from track properties
  %for tracks identified as pion and electron/positron 
  for
  MC-simulated 
  $\Ks \Kl \to (\pi^{\pm}e^{\mp}\nu)(3\pi^0)$
  %$\Ks\to\pi^{\pm}e^{\mp}\nu$ 
  events (left) and all data events (right).
  Events inside the region marked with black solid line are accepted.}
\label{fig:t1-tof3}
\end{figure}

After the pion and electron tracks are identified, a second cut is applied
in the $\delta t(\pi)$ \textit{vs} $\delta t(e)$ plane as shown in Fig.~\ref{fig:t1-tof3}
to further discriminate the remaining $\pi^+\pi^-$ events.
After these cuts, background remains (amounting to about 12\% of the event sample) constituted by the $\Ks\to\pi^0\pi^0$ and $\Kl\to\pi e\nu$ processes
where the $\Ks$ decay along with additional clusters had been misidentified as an early $\Kl\to 3\pi^0$ decay.
This background is discriminated by removing events containing more than one cluster for
which $R/(cT_{clu}) > 0.9$ where $R$ is the distance between the 
%$e^+e^-$ interaction point~(IP) 
IP
and the cluster~(corresponding to photons emitted close to the IP).

The remaining background (in the order of decreasing contribution) is composed of:
\begin{itemize}
\item $\Ks\to\pi^+\pi^-$ with imperfect track reconstruction,
\item $\Ks\to\pi^+\pi^-\to \pi\mu\nu$ decay chain where one of the charged pions decays into a muon and a neutrino before entering the DC,
\item radiative $\Ks\to\pi^+\pi^-\gamma$ decays dominated by inner bremsstrahlung~\cite{PhysRevLett.70.2525,Gatti:2005kw}.
  % where the photon is radiated by one of the pions and carries away a part of its momentum.
\end{itemize}
As all these events are characterised by a pion or muon track misidentified as $e^+/e^-$,
two particle binary classifiers based on Artificial Neural Networks~(ANNs)
(using Multilayer Perceptron from the TMVA package~\cite{Hocker:2007ht})
and acting on an ensemble constituted by a track and its associated cluster
are prepared for $e/\pi$ and $\mu/\pi$ discrimination in subsequent steps.
Classification is based on the different structure of 
the energy deposited in the EMC cells 
%electromagnetic showers caused in the EMC 
by electrons, pions, and muons, in combination with the information of the
%difference as well as the total deposited energy and 
particle momentum from the associated DC track.
The ANNs are trained using data control samples of $\Kl\to\pi^{\pm}e^{\mp}\nu$ and $\Kl\to\pi^{\pm}\mu^{\mp}\nu$
decays tagged by $\Ks\to\pi^+\pi^-$ identified with a 98\% purity according to MC simulations.

\begin{figure}[h!]
  \centering
  \includegraphics[width=0.8\linewidth]{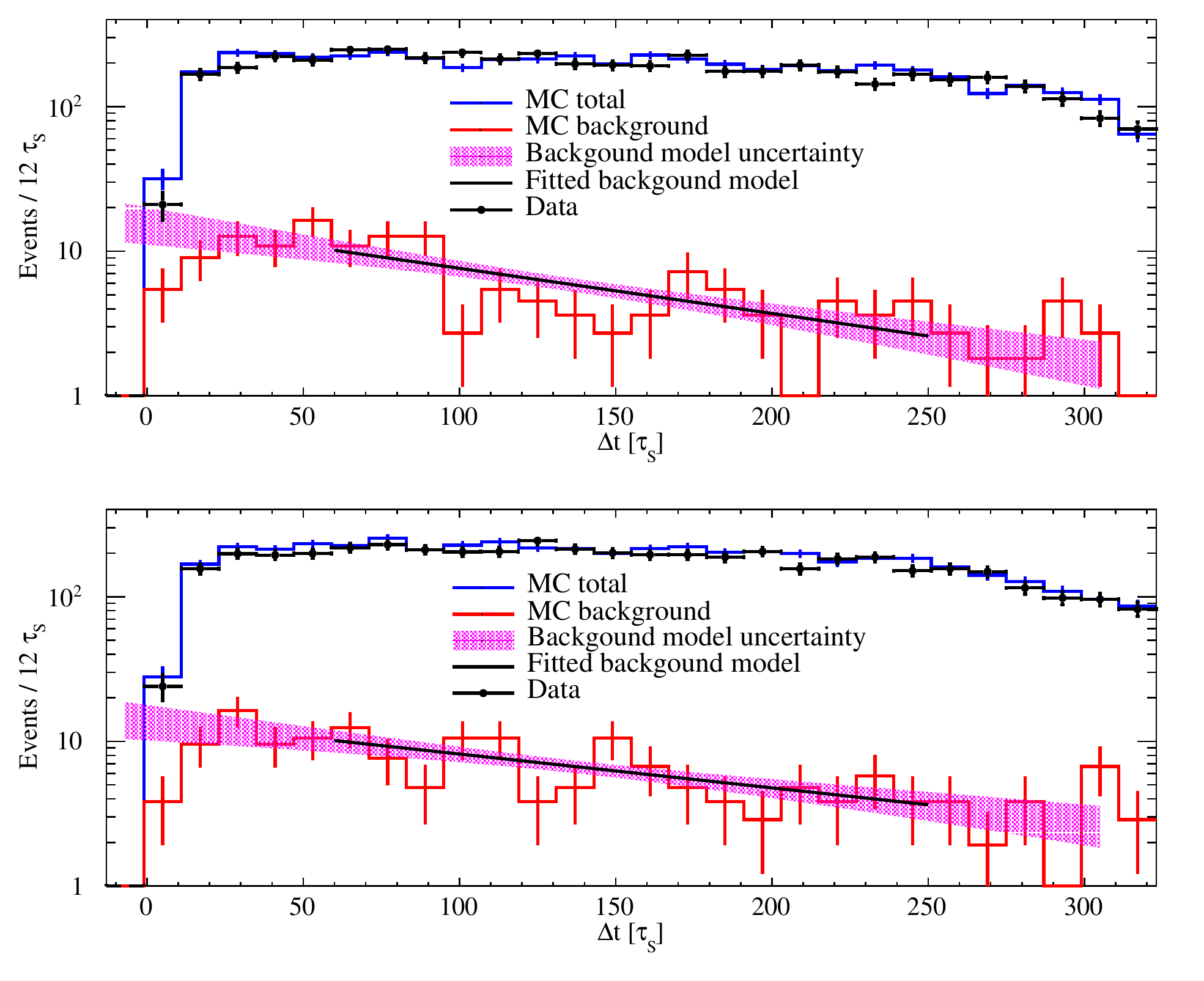}
  \caption{%
    $\Ks\Kl\to(\pi^{\pm}e^{\mp}\nu)(3\pi^0)$ sample:
    exponential model of residual background distribution as a function of $\Delta t$ 
  %time difference between kaon proper decay times
    for events with an electron (top) and positron (bottom) in the early kaon semileptonic decay.
    The modelled expected background is obtained in a fit to MC residual background
    and subtracted from the respective data distributions. Uncertainty of the fitted models (magenta band)
    is taken into account as a source of systematic uncertainty.
  }\label{fig:bcg-model}
\end{figure}

After the $e/\pi$ and $e/\mu$ particle discrimination for lepton track candidates,
the signal to background ratio
amounts to 22.5 with residual background dominated by $\Ks\to\pi^+\pi^-$~(55\%), $\Ks\to\pi^+\pi^-\gamma$~(19\%)
and events with $\Kl$ decays other than $3\pi^0$~(12\%).
%As the purity of the event sample was almost three times lower
%than in case of the $\Ks \Kl \to (\pi^{+}\pi^{-})(\pi^{\pm}e^{\mp}\nu)$ process,
Distribution of the residual background as a function of
$\Delta t$
is modelled with an exponential function presented in Fig.~\ref{fig:bcg-model}.
Parameters of the exponential models are obtained with a fit to MC residual background separately  
for event samples with an electron and positron with $\chi^2$/NDF of 2.1 and 0.99 respectively.
These models are used to subtract the expected background 
contamination from the data distributions shown in black points in Fig.~\ref{fig:bcg-model}.

\subsection{Selection of $\Ks \Kl \to (\pi^{+}\pi^{-})(\pi^{\pm}e^{\mp}\nu)$ events}\label{sec:selection2}
As $\Km\to\kaon$ and $\Km\to\akaon$ transitions are characterised by an early kaon decay into two pions
(the $\pi^+\pi^-$ final state is chosen in this analysis) followed by a later semileptonic decay,
event selection requires presence of a DC vertex associated to 2 opposite curvature tracks within a cylindrical volume of 
$\rho < 15$~cm and $|z|<10$~cm around the IP and with
$|m_{\pi\pi} - m_{\kaon}| < 10$~MeV/c\textsuperscript{2}.

\begin{figure}[h!]
  \centering
  \includegraphics[width=0.5\linewidth]{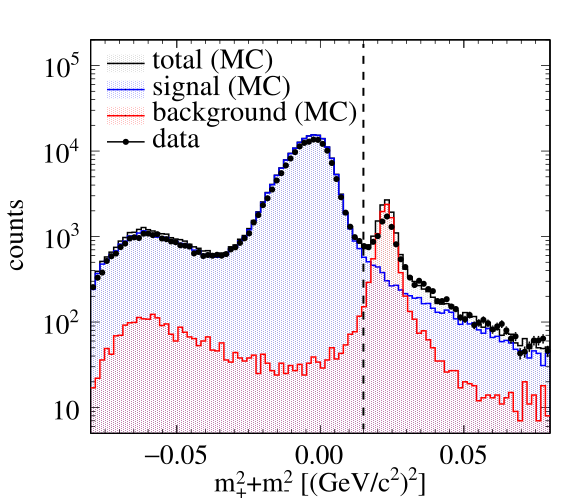}
  \caption{Sum of invariant masses of two decay product tracks assuming their electron mass hypotheses and a semileptonic decay.
    Background and data distributions contain contributions from different decay vertex choices. $\Kl$ decay vertex candidates with
    $m^2_{+} + m^2_{-} > 0.015\:(\mathrm{GeV}/c^{2})^2$ (i.e.\ right of the dashed vertical line) are rejected.}\label{fig:t2-m2}
\end{figure}

To find the semileptonic decay vertex, all vertices formed by two opposite curvature tracks in the DC
are considered,
excluding the previously identified $\Ks\to\pi^+\pi^-$ vertex and its related tracks.
%For each of the vertex candidates and each of the tracks (referred to by the particle charge),
%the invariant mass of the associated particle assuming electron mass was evaluated:
For each candidate semileptonic vertex the following invariant mass
(which for correct $e/\pi$ identification should correspond to the electron mass)
is evaluated separately for each track:

\begin{equation}
  m^2_{\pm} = (E_K-E(\pi)_{\mp}-|\vec{p}_{miss}|)^2 - |\vec{p}_{\pm}|^2,
\end{equation}
with $E_K$ being the energy of the decaying kaon, $E(\pi)_{\mp}$ 
%-- 
the energy attributed to the track of negative (positive) charge
%with 
%a
assuming the charged pion mass, 
% the assumption it was a charged pion
%is 
$\vec{p}_{\pm}$~the momentum corresponding to positive (negative) charge track,
and $\vec{p}_{miss} = \vec{p}_{K} - \vec{p}_{+} - \vec{p}_{-}$. 
In the dominant background sources ($\pi^+\pi^-\pi^0$ and semileptonic decays with a muon)
both tracks are characterised by significant $m^2$ values whereas for genuine $\pi^{\pm}e^{\mp}\nu$ decays
$m^2\approx 0$ for one of the tracks, making the sum of $m^{2}$ for both tracks in the event sensitive to
a difference between $\Kl\to\pi^{\pm}e^{\mp}\nu$ vertices and other decays as shown in Fig.~\ref{fig:t2-m2}.
Vertex candidates with $m^2_{+} + m^2_{-} > 0.015\:(\mathrm{GeV}/c^{2})^2$
are rejected and only events with one remaining vertex candidate are considered in further analysis.

Further selection of semileptonic decays of $\Kl$ as well as identification
of the $e^{\pm}$ and $\pi^{\pm}$ tracks is performed with a time of flight
analysis similarly as in the case of $\Ks\to\pi^{\pm}e^{\mp}\nu$,
resulting in signal to background ratio of 75.
The remaining background is neglected in further analysis.

\subsection{Determination of efficiencies}\label{sec:efficiencies}
For each of the two classes of processes, selected events are split by charge of the electron and positron
from the semileptonic decay.
$\Delta t$-dependent efficiencies are evaluated separately for each of the obtained four samples of events as:
\begin{equation}
  \varepsilon_{total}(\Delta t) =  \varepsilon_{TEC} \times \varepsilon_{SEL}(\Delta t),
\end{equation}
where $\varepsilon_{TEC}$ is the combination of $\Delta t$-independent efficiencies
of trigger, machine background filter and event classification
described in Section~\ref{sec:kloe}
and $\varepsilon_{SEL}(\Delta t)$ represents the efficiency of event selection (cuts) for a particular $\Delta t$~interval.

For evaluation of the trigger efficiency, rates of events with either one -- EMC or DC based -- or both triggers
for events passing the entire event selection, are used to obtain the total probability of at least one of the triggers
present on a signal event.
Efficiency of the machine background filter is estimated by counting
events passing the signal selection criteria among
10\% of all events acquired independently of the filter decision.
Similarly, a subsample of 5\% of all events stored independently of their classification is used to evaluate the
probability of misclassification of a signal event and thus the classification efficiency.
Table~\ref{tab:e-tec} summarises the $\varepsilon_{TEC}$ efficiencies combining these three factors
for each class of signal events.

\begin{table}[h!]
  \centering
    \caption{Combined efficiencies of trigger, machine background filter and event classification for the four event samples.
  \label{tab:e-tec}}
  \begin{tabular}{cc}
    event sample & $\varepsilon_{TEC}$~[\%] \\ \hline
    $(\pi^+e^-\nu) \; (3\pi^0)$ & 99.487 $\pm$ 0.071 \\
    $(\pi^-e^+\nu) \; (3\pi^0)$ & 99.453 $\pm$ 0.074 \\
    \hline
    $(\pi^+\pi^-) \; (\pi^+e^-\nu)$ & 99.597 $\pm$ 0.007 \\
    $(\pi^+\pi^-) \; (\pi^-e^+\nu)$ & 99.589 $\pm$ 0.010 \\
  \end{tabular}
\end{table}

Event selection efficiencies are estimated using Monte Carlo simulations of the respective event classes.
In case of semileptonic decays, MC-based efficiencies $\varepsilon_{SEL}(\Delta t)$ are 
corrected 
%for the DATA/MC ratio 
%optimized 
using an independent data control sample of
$\Ks \Kl \to (\pi^{0}\pi^{0})(\pi^{\pm}e^{\mp}\nu)$ events. The selection efficiency for $\Kl\to 3\pi^0$~decays
is estimated with MC and corrected
%optimized 
with a $\Ks \Kl \to (\pi^{+}\pi^{-})(3\pi^0)$ data control sample.

%%%%%%%%%%%%%%%%%%%%%%%%%%%%%%%%%%%%%%%%%%%%%%%%%%%%%%%%%%%%%%%%%%%%%%%%%%
% Construction of ratios                                                 %
%%%%%%%%%%%%%%%%%%%%%%%%%%%%%%%%%%%%%%%%%%%%%%%%%%%%%%%%%%%%%%%%%%%%%%%%%%
\section{Ratios of double kaon decay rates}\label{sec:ratios}
\begin{figure*}[h!]
  \centering
  \includegraphics[width=0.8\linewidth]{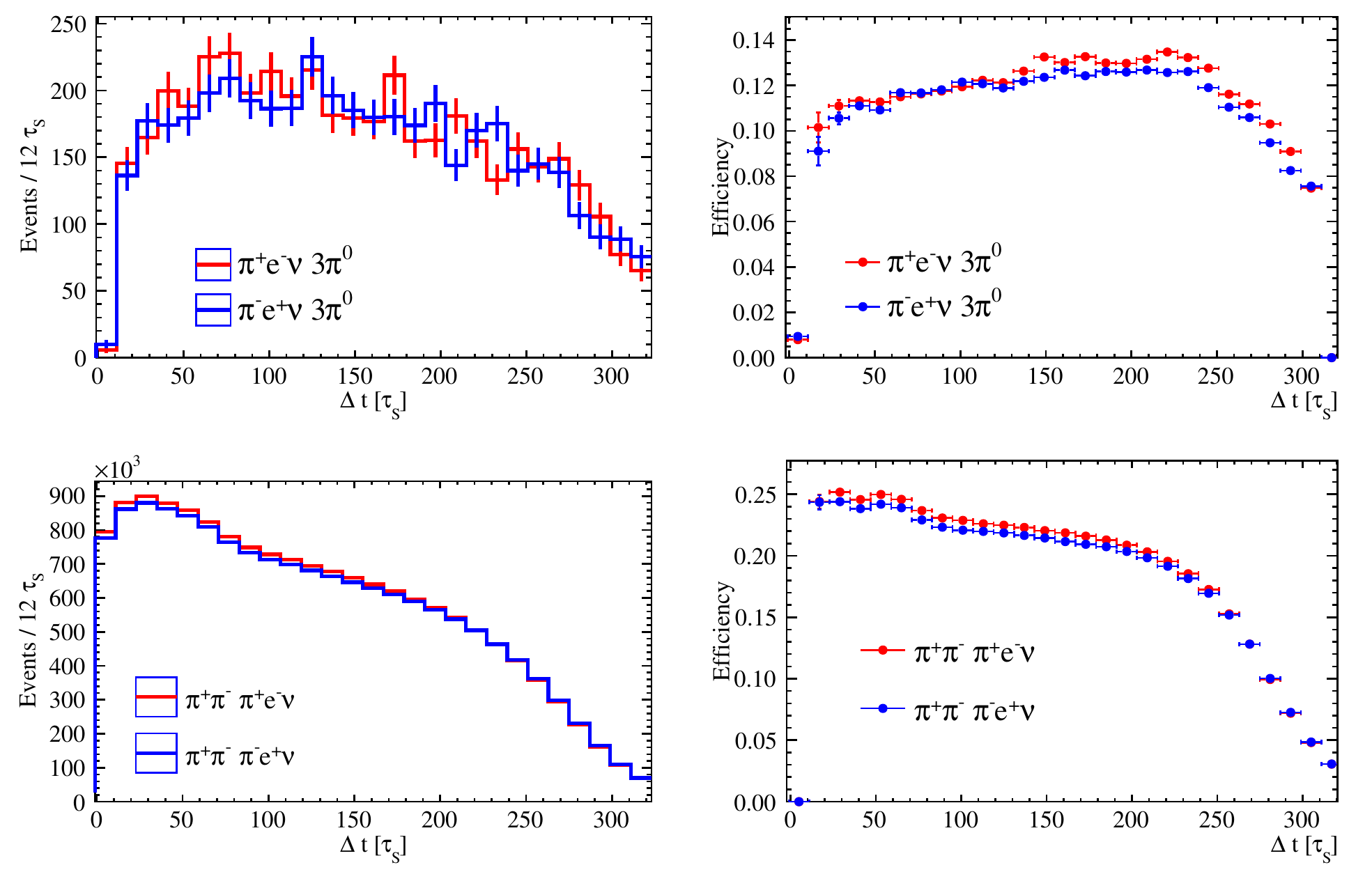}
  \caption{Left column: Rates of double kaon decays as a function of $\Delta t$ 
  %the difference of kaon proper decay times 
  for the two studied classes of processes and two lepton charge subsamples. Right column: corresponding 
  $\varepsilon_{SEL}(\Delta t)$
  %total event identification 
  efficiencies.}
  \label{fig:dt_plots}
\end{figure*}

After event selection, the counts of events identified
for each class are presented in Table~\ref{tab:event-counts}.
Fig.~\ref{fig:dt_plots} shows a summary of the corresponding
data $\Delta t$ distributions %for the four charge subsamples
%of two event classes
entering the probability ratios along with their respective $\varepsilon_{SEL}(\Delta t)$
%total identification
event selection
efficiencies in the range $0 < \Delta t < 320\:\tau_S$,
with a bin width of $12\:\tau_S$.
%obtained as described in Section~\ref{sec:efficiencies}.
%Data was binned into $\Delta t$ intervals of 12$\tau_{S}$
%in order to maintain over 100 events per bin in all distributions.

\begin{table}[h!]
  \centering
  \caption{Numbers of events identified for each of the four event classes.\label{tab:event-counts}}
  \begin{tabular}{cc}
    event sample & events \\ \hline
    $(\pi^+e^-\nu) \; (3\pi^0)$ & 4750\\
    $(\pi^-e^+\nu) \; (3\pi^0)$ & 4652 \\
    \hline
    $(\pi^+\pi^-) \; (\pi^+e^-\nu)$ & 15924863 \\
    $(\pi^+\pi^-) \; (\pi^-e^+\nu)$ & 15708190 \\
  \end{tabular}
\end{table}

\begin{figure*}[h!]
  \centering
  \includegraphics[width=0.6\linewidth]{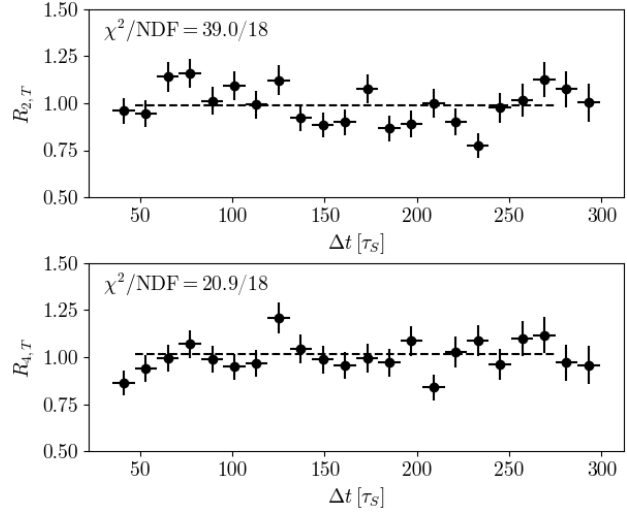}
  \caption{T-violation sensitive ratios of double decay rates 
  %of entangled $\kaon\akaon$ pairs
    as defined in Eq.~(\ref{ratio2t}).
    Dashed lines denote levels obtained with the fit.
  }
  \label{fig:t-ratios}
\end{figure*}

\begin{figure*}[h!]
  \centering
  \includegraphics[width=0.6\linewidth]{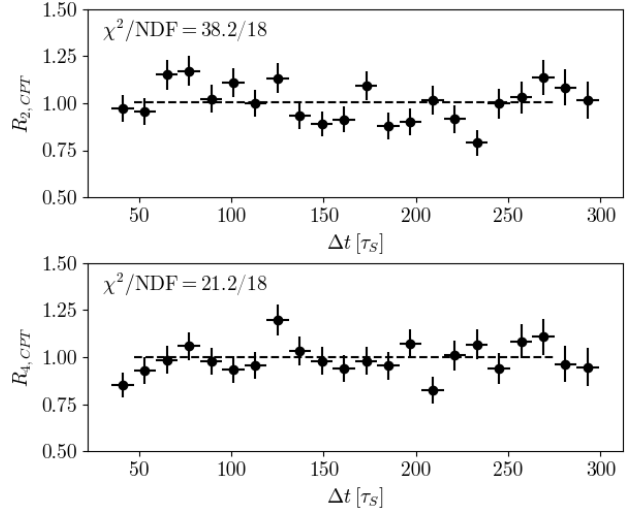}
  \caption{CPT-violation sensitive ratios of double decay rates 
  %of entangled $\kaon\akaon$ pairs
    as defined in Eq.~(\ref{ratio2cpt}).
    Dashed lines denote levels obtained with the fit.
  }
  \label{fig:cpt-ratios}
\end{figure*}

The T and CPT-violation sensitive single ratios defined 
in Eqs.~(\ref{ratio4t}) and (\ref{ratio4cpt})
are shown in Figs.~\ref{fig:t-ratios} and~\ref{fig:cpt-ratios}.
%are evaluated in subsequent intervals of the difference of kaons' proper decay times $\Delta t$.
Each point of the single ratio graphs
is defined through the counts of the respective double kaon decays
$N_i$ and $N'_i$ in the $i$-th interval of $\Delta t$
and their corresponding event identification efficiencies $\varepsilon_i$ and $\varepsilon'_i$ as:
\begin{equation}
  R_i \equiv R(\Delta t_i) = \frac{N_i}{N_i'}\frac{\varepsilon_i'}{\varepsilon_i} \frac{1}{D},
\end{equation}
where $D$ is the factor defined in Section~\ref{sec:introduction}.

Due to the limited statistics of the process entering the numerator
of the ratios, a constant level of the single ratios
is evaluated in the range of high and relatively stable efficiency of
$\Delta t \in (47,275)\tau_{S}$
using a maximum likelihood fit for the constant ratio $R_{\SS}$:
%minimizing :
\begin{equation}
  \label{eq:fit_likelihood}
  \mathcal{L}(R_{\SS}) = \prod_{i \text{ in fit limits}} p\left( N_i,\: 
  R_{\SS}
  %r 
  N_i'D\frac{\varepsilon_i}{\varepsilon_i'} \right),
\end{equation}
where $p(n,\:\lambda)$ is the Poissonian probability of observing $n$
counts with the distribution mean of $\lambda$.

\begin{figure}[h!]
  \centering
  \includegraphics[width=0.6\linewidth]{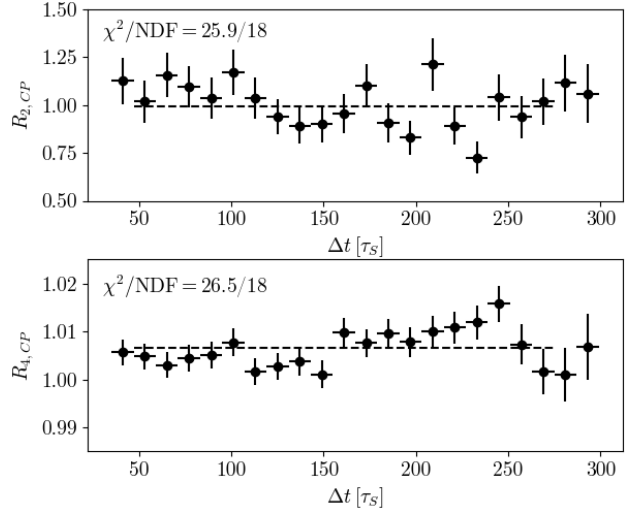}
  \caption{Ratios of the rates of
    $(\pi^{\pm}e^{\mp}\nu)(3\pi^0)$~(top) and
    $(\pi^{+}\pi^{-})(\pi^{\pm}e^{\mp}\nu)$~(bottom)
    events sensitive to CP-violation effects.
    Dashed lines denote levels obtained with the fit.
  }
  \label{fig:cp-ratios}
\end{figure}

\begin{figure}[h!]
  \centering
  \includegraphics[width=0.6\linewidth]{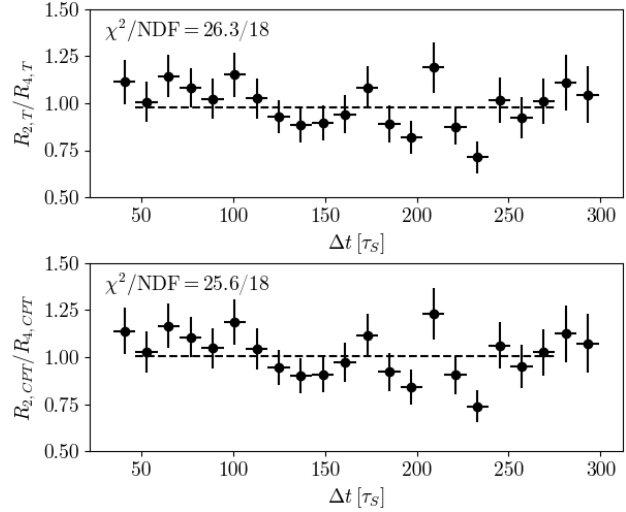}
  \caption{Double ratios of double kaon decay rates
    sensitive to effects of T and CP~violation~(top) and
    CPT~violation~(bottom).
    Dashed lines denote levels obtained with the fit.
  }
  \label{fig:double-ratios}
\end{figure}

Counts of the $(\pi^{+}\pi^{-})(\pi^{\pm}e^{\mp}\nu)$ events with a positron and an electron
are also used to construct transition rate ratios sensitive to CP-violation
(Eq.~(\ref{ratio4cp}))
shown in Fig.~\ref{fig:cp-ratios}.
Finally, double ratios sensitive to T and CP violation (Eq.~(\ref{eq:ratiosdef7})) and CPT violation (Eq.~(\ref{eq:ratiosdef8}))
are constructed and their asymptotic levels are estimated as presented in Fig.~\ref{fig:double-ratios}.

%%%%%%%%%%%%%%%%%%%%%%%%%%%%%%%%%%%%%%%%%%%%%%%%%%%%%%%%%%%%%%%%%%%%%%%%%%
% Systematics                                                            %
%%%%%%%%%%%%%%%%%%%%%%%%%%%%%%%%%%%%%%%%%%%%%%%%%%%%%%%%%%%%%%%%%%%%%%%%%%
\section{Systematic uncertainties}\label{sec:systematics}
Stability of the results is checked 
%by
by varying the event selection cut values by $\pm 5 \sigma$ in steps of $1~\sigma$
where $\sigma$ denotes the resolution on the variable subject to the cut,
repeating the analysis for each cut value
and observing the impact of variation on each of the eight observables of the tests.
For evaluation of the corresponding systematic uncertainties
%arising from chosen cut values,
absolute deviations of each of the observables for $\pm 1 \sigma$ cut variation
are averaged.

The uncertainty due to the model of subtracted background
is estimated by varying the two model parameters within their errors.
%Impact of the uncertainty of the subtracted model of residual background (Figure~\ref{fig:bcg-model}.) 
%was estimated as variation of the observable results obtained
%including subtraction of a model when one of the parameters was varied within its error.
Effects from $\pm 1 \sigma$ variations of each parameter of the model are added in quadrature.

Event selection efficiencies based on MC simulations with limited statistics
are subject to a smoothing procedure. The corresponding systematic uncertainty is quantified
by comparing analysis results with and without efficiency smoothing.

The choice of $\Delta t$ bin width for the final ratio distributions and the fit limits
account for another source of systematic uncertainty. Stability of the fit results
is tested for bin widths ranging from 3~$\tau_S$ to 24~$\tau_S$
and an average of the observables' deviations obtained with the extreme bin widths
from the result with 12~$\tau_{S}$ is used as an estimate of the systematic effect.
The fitting range is varied by $\pm24\ \tau_{S}$ in steps of $\pm 6\ \tau_{S}$ as a shift as well as total width.
Changes of the fitted ratio levels for variations of $\pm 6\ \tau_{S}$ are averaged and added in quadrature.

Table~\ref{tab:systematics} summarises all identified systematic effects
on each of the eight observables of the tests.
%Contributions from subsequent event selection cuts and other parameters
%of ratio levels' extraction 
All contributions
are added in quadrature to obtain the total systematic uncertainties indicated in bold.
In case of single ratios, the results are additionally affected by the total error
on the $D$ factor
(quoted separately due to containing both statistical and systematic contributions)
obtained from previous KLOE measurements.
%As these uncertainties comprise a combination of statistical and systematic errors, they are quoted separately.

\begin{table*}[h!]
\caption{Systematic uncertainties on all of the symmetry test observables.
  In case of $R_{2,CP}$ and $R_{4,CP}$, each ratio is obtained using only one class of events
  and thus not affected by effects of selection of the other class.
  The uncertainty of the D factor comprises both statistical and systematic errors.
  \label{tab:systematics}}

  \begin{adjustwidth}{-0.15\columnwidth}{}
  \centering
  \resizebox{1.3\columnwidth}{!}{%
    \input{syst_table.tex}

  }
  \end{adjustwidth}
\end{table*}

%%%%%%%%%%%%%%%%%%%%%%%%%%%%%%%%%%%%%%%%%%%%%%%%%%%%%%%%%%%%%%%%%%%%%%%%%%
% Results                                                                %
%%%%%%%%%%%%%%%%%%%%%%%%%%%%%%%%%%%%%%%%%%%%%%%%%%%%%%%%%%%%%%%%%%%%%%%%%%
% \section{Results}\label{sec:results}
% Constant levels of the single ratios defined in Eqs.~\ref{ratio2t}--\ref{ratio4cpt}
% and two double ratios of double kaon decay rates (Eqs. \ref{eq:ratiosdef7}--\ref{eq:ratiosdef8})
% were evaluated with a fit to the distributions presented in Figs.~\ref{fig:single-ratios}--\ref{fig:double-ratios}.
% The following levels of the ratios were obtained:

% No symmetry-violating effects were observed within the uncertainty of
% the observables except for $R^{exp}_{4,CP}$ which,
% unlike other ratios, is not limited in sensitivity
% by the small branching fraction of semileptonic $\Ks$ decays.
% Effects of CP violation are visible as an offset of this ratio level over unity.
% with magnitude consistent with previous measurements of $\Re(\epsilon)$~\cite{pdg}
% within total uncertainty.

%% file: syst_table.tex
\begin{tabular}{p{6.0cm}rrrrrrrr}
\hline
  Effect &  $R_{2,T}$ & $R_{4,T}$ &  $R_{2,{CPT}}$ & $R_{4,{CPT}}$ & $DR_{T,CP}$ & $DR_{CPT}$ & $R_{2,CP}$ & $R_{4,CP}$  \\
% Effect &  $R_{2,T}$ & $R_{4,T}$ &  $R_{2,{CPT}}$ & $R_{4,{CPT}}$ & $R_{2,T}/R_{4,T}$ & $\frac{R_{2,{CPT}}}{R_{4,{CPT}}}$ & $R_{2,CP}$ & $R_{4,CP}$  \\
  & $\times\ 10^{-3}$ & $\times\ 10^{-3}$ & $\times\ 10^{-3}$ & $\times\ 10^{-3}$ & $\times\ 10^{-3}$ & $\times\ 10^{-3}$ & $\times\ 10^{-3}$ & $\times\ 10^{-3}$ \\
  \hline
  Background model &  2.74 &  4.62 &  2.79 &  4.43 &  4.43 &  4.41 &  4.37 &  -- \\
  Efficiency smoothing &  2.46 &  5.31 &  2.43 &  5.26 &  6.70 &  6.83 &  6.76 &  0.17 \\
  $\Delta t$ bin width &  8.00 &  5.00 &  7.50 &  5.50 &  9.00 &  9.00 &  8.90 &  0.03 \\
  Fit range & 7.33 & 8.88 & 7.32 & 8.84 & 7.95 & 7.60 & 7.78 & 0.41 \\
  \hline
  \multicolumn{9}{l}{Effects of cuts in the $(\pi e \nu)(3\pi^0)$ selection}\\ 
%  \hline
  $\Ks$ vertex location cuts & 0.57 & 2.31 & 0.58 & 2.27 & 2.36 & 2.41 & 2.39 & -- \\
  M($\pi,\pi$) cut &  2.48 &  1.34 &  2.52 &  1.31 &  1.56 &  1.63 &  1.60 &  -- \\
  TOF cuts & 6.08 & 5.32 & 6.19 & 5.23 & 6.40 & 6.58 & 6.49 & -- \\
  e/$\pi$/$\mu$ classification & 4.78 & 4.40 & 4.85 & 4.33 & 9.33 & 9.59 & 9.46 & -- \\
  \hline
  \multicolumn{9}{l}{Effects of cuts in the $(\pi^+\pi^-)(\pi e \nu)$ selection}\\  
%  \hline
  $\Ks$ vertex location cuts & 0.007 & 0.004 & 0.004 & 0.007 & 0.004 & 0.004 & -- & 0.005\\
   M($\pi,\pi$) and $|\vec{p}|$ cuts & 2.14 & 1.68 & 1.67 & 2.17 & 0.70 & 0.72 & -- & 0.74 \\
  $m^2_{+} + m^2_{-}$ cut &  1.48 &  1.32 &  1.31 &  1.49 &  0.20 &  0.21 &  -- &  0.21 \\
  TOF cuts & 2.14 & 1.68 & 1.67 & 2.17 & 0.70 & 0.72 & -- & 0.74\\
  \hline
  \textbf{Total systematic \mbox{uncertainty}} &  \textbf{14} &  \textbf{15} &  \textbf{14} &  \textbf{15} &  \textbf{19} &  \textbf{19} &  \textbf{19} &  \textbf{0.89} \\
  D factor total uncertainty &  12 &  12 &  12 &  12 & -- & -- & -- & -- \\
\hline
\end{tabular}

%% file: conclusions.tex
\section{Results and conclusions \label{sec:conclusions}}
\par
We presented the first direct tests 
of \T, \CP, \CPT  symmetries in transitions of neutral kaons,
obtained analysing $1.7~\hbox{fb}^{-1}$ of data collected by the KLOE experiment at
DA$\Phi$NE.
The \T and \CPT tests involving time-reversal  implement
the necessary exchange of 
{\it in} and {\it out} states, as required for a genuine test, 
exploiting the entanglement of the $\kn\knb$ pairs.
%to realize the necessary exchange of 
%{\it in} and {\it out} states 
The decay intensities of the processes 
$\phi\rightarrow\ksn\kln\rightarrow\pi^+\pi^- \, \pi e \nu $ and 
$\phi\rightarrow\ksn\kln\rightarrow \pi e \nu \, 3\pi^0$
as a function of $\Delta t$ are measured in the asymptotic region $\Delta t \gg \tau_S$, statistically the most significant.
We get results for all eight observables defined in Eqs.~(\ref{eq:ratiosdef1})--(\ref{eq:ratiosdef8}):
\begin{eqnarray*}
  R_{2,T} = & 0.991 &\pm 0.017_{stat} \pm 0.014_{syst} \pm {0.012}_D,\\
  R_{4,T} = & 1.015 &\pm 0.018_{stat} \pm 0.015_{syst} \pm {0.012}_D,\\
  R_{2,{CPT}} =& 1.004 &\pm 0.017_{stat} \pm 0.014_{syst} \pm {0.012}_D,\\
  R_{4,{CPT}} =& 1.002 &\pm 0.017_{stat} \pm 0.015_{syst} \pm {0.012}_D,\\
  R_{2,{CP}} =& 0.992 &\pm 0.028_{stat} \pm 0.019_{syst},\\
  R_{4,{CP}} =& 1.00665 &\pm 0.00093_{stat} \pm 0.00089_{syst},\\
  DR_{T,CP} = R_{2,{T}} / R_{4,{T}} =& 0.979 &\pm 0.028_{stat} \pm 0.019_{syst},\\
  DR_{CPT} = R_{2,{CPT}} / R_{4,CPT} =& 1.005 &\pm 0.029_{stat} \pm 0.019_{syst}.
\end{eqnarray*}

\par
\begin{figure}[h!]
  \centering
  \includegraphics[width=0.7\linewidth]{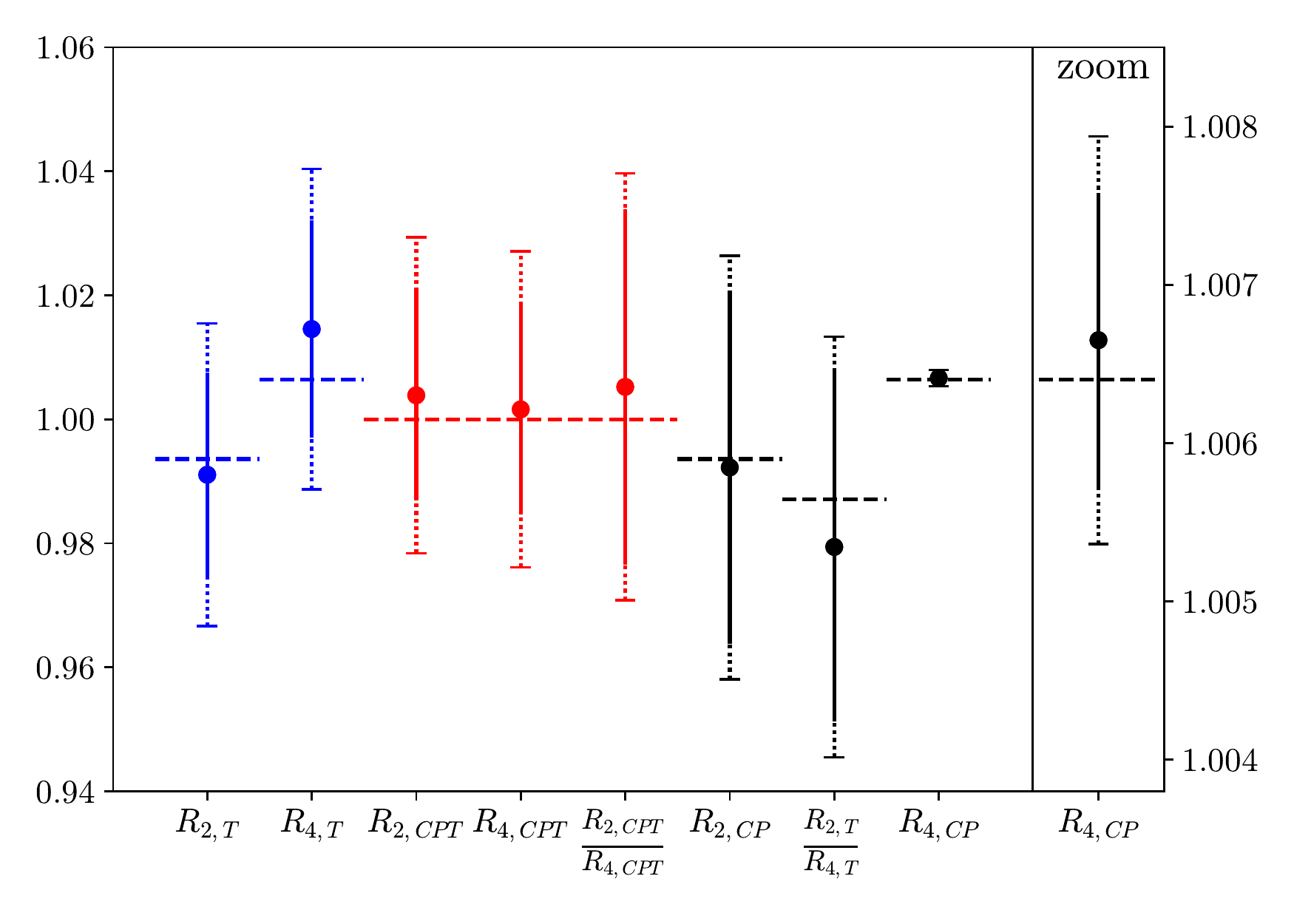}
  \caption{
    Comparison of the measured symmetry-violation-sensitive single and double ratios and their expected values
    (horizontal dashed lines) assuming \CPT invariance and \T violation extrapolated from observed \CP violation in the mixing. Solid error bars denote statistical uncertainties and dotted error bars represent total uncertainties (including error on the D factor in case of single \T and \CPT-violation sensitive ratios).
    The right-hand-side panel magnifies the region of the \CP-violation-sensitive ratio $R_{4,CP}.$
  }\label{fig:final-comparison}
\end{figure}

A comparison of these results
%measured levels 
with expected values %(see Eqs. \ref{eq:ratiosdef1}--\ref{eq:ratiosdef8}) 
 (assuming \CPT invariance and \T violation extrapolated from observed \CP violation in the mixing)
is presented in Fig.~\ref{fig:final-comparison}.

\par
For the \T and \CPT single ratios 
a total relative error of $2.5~\%$ is reached, while for the double ratios (\ref{eq:ratiosdef7}) and (\ref{eq:ratiosdef8})
%, that are independent on the $D$ factor
%connecting the experimental
%intensities with the transition probabilities, 
%not depending on the $D$ factor,
the total error increases
to 
%of $\sqrt{2}$, 
%(i.e. 
$3.5~\%$,
%), 
with 
%in principle 
the advantage of %in principle a
%doubled
improved
sensitivity to violation effects, and of independence from the $D$ factor.
The measurement of the single ratio $R_{4,\CP}$ benefits of highly allowed decay rates for the involved channels, reaching an error of $0.13~\%$.
%a sensitivity to violation effects doubled with respect to single ratios.
\par 
The double ratio $DR_{\CPT}$
is our best observable for testing \CPT,
%one of the cleanest ever performed.
free from approximations and model independent, while $DR_{\T,\CP}$ assumes no direct \CPT violation and is 
%even under \CPT -- contrary to single ratios -- and therefore it is not useful to disentangle \T from \CP violation effects.
even under \CPT, 
therefore it does not disentangle \T and \CP violation effects, contrary to the 
genuine \T and \CP single ratios.
%bona-fide
%TRV and CPV single ratios.
%therefore not useful to disentangle \T from \CP violation effects, contrary to single \T and \CP single ratios.
\par
No result on \T and \CPT observables shows evidence of symmetry violation.
We observe \CP violation in transitions in the single ratio 
$R_{4,\CP}$
%(\ref{eq:ratiosdef4}) 
with a significance of $5.2\sigma$, in agreement with
the known \CP violation in the $\kn-\knb$ mixing using a different observable.
%in the measurement of the charge semileptonic asymmetry of \kln~\cite{ktev}, 
%though $R_{4,\CP}$ being a different observable connected with $\knn \rightarrow \kn,\knb$ transitions.